\UseRawInputEncoding % KP

\documentclass[aps,prc,twocolumn,groupedaddress,footnote,thanks]{revtex4-2}

\usepackage{amsmath}
\usepackage{amssymb}

\usepackage{graphicx}% Include figure files
\usepackage{dcolumn}% Align table columns on decimal point
\usepackage{bm}% bold math
\begin{document}

%\preprint{}

\title{
Production of hydrogen isotopes and charged pions in p (3.5 GeV) + Nb reactions 
}


\author{
R.~Abou~Yassine$^{6,14}$, O.~Arnold$^{10,9}$, M.~Becker$^{11}$, P.~Bergmann$^{5}$,
A.~Blanco$^{1}$, C.~Blume$^{8}$, M.~B\"{o}hmer$^{10}$, N.~Carolino$^{1}$, L.~Chlad$^{15,c}$,
P.~Chudoba$^{15}$, I.~Ciepa{\l}$^{3}$, J.~Dreyer$^{7}$, W.A.~Esmail$^{5}$, L.~Fabbietti$^{10,9}$,
P.~Fonte$^{1,a}$, J.~Friese$^{10}$, I.~Fr\"{o}hlich$^{8}$, T.~Galatyuk$^{6,5}$, J.~A.~Garz\'{o}n$^{16}$,
M.~Grunwald$^{17}$, M.~Gumberidze$^{5}$, S.~Harabasz$^{6}$, C.~H\"{o}hne$^{11,5}$, F.~Hojeij$^{14}$,
R.~Holzmann$^{5}$, H.~Huck$^{8}$, M.~Idzik$^{2}$, B.~K\"{a}mpfer$^{7,b}$, B.~Kardan$^{8}$,
V.~Kedych$^{6}$, I.~Koenig$^{5}$, W.~Koenig$^{5}$, M.~Kohls$^{8}$, G.~Korcyl$^{4}$,
G.~Kornakov$^{17}$, F.~Kornas$^{6,5}$, R.~Kotte$^{7}$, W.~Krueger$^{6}$, A.~Kugler$^{15}$,
T.~Kunz$^{10}$, R.~Lalik$^{4}$, L.~Lopes$^{1}$, M.~Lorenz$^{8}$, A.~Malige$^{4}$,
J.~Markert$^{5}$, V.~Metag$^{11}$, J.~Michel$^{8}$, A.~Molenda$^{2}$, C.~M\"{u}ntz$^{8}$,
L.~Naumann$^{7}$, K.~Nowakowski$^{4}$, J.-H.~Otto$^{11}$, Y.~Parpottas$^{12}$, M.~~Parschau$^{8}$,
V.~Pechenov$^{5}$, O.~Pechenova$^{5}$, J.~Pietraszko$^{5}$, A.~Prozorov$^{15,c}$, W.~Przygoda$^{4}$,
K.~Pysz$^{3}$, B.~Ramstein$^{14}$, N.~Rathod$^{4}$, A.~Rost$^{6,5}$, A.~Rustamov$^{5}$,
P.~Salabura$^{4}$, N.~Schild$^{6}$, E.~Schwab$^{5}$, F.~Seck$^{6}$, U.~Singh$^{4}$,
S.~Spies$^{8}$, M.~Stefaniak$^{17,5}$, H.~Str\"{o}bele$^{8}$, J.~Stroth$^{8,5}$, C.~Sturm$^{5}$,
K.~Sumara$^{4}$, O.~Svoboda$^{15}$, M.~Szala$^{8}$, P.~Tlusty$^{15}$, M.~Traxler$^{5}$,
H.~Tsertos$^{13}$, V.~Wagner$^{15}$, A.A.~Weber$^{11}$, C.~Wendisch$^{5}$, H.P.~Zbroszczyk$^{17}$,
E.~Zherebtsova$^{5,d}$, P.~Zumbruch$^{5}$\\
(HADES collaboration) \\
and B.~Kamys$^{4}$, S.~Sharma$^{4}$\\
}
\affiliation{
\ \\
\mbox{$^{1}$LIP-Laborat\'{o}rio de Instrumenta\c{c}\~{a}o e F\'{\i}sica Experimental de Part\'{\i}culas , 3004-516~Coimbra, Portugal}\\
\mbox{$^{2}$AGH University of Science and Technology, Faculty of Physics and Applied Computer Science, 30-059~Kraków, Poland}\\
\mbox{$^{3}$Institute of Nuclear Physics, Polish Academy of Sciences, 31342~Krak\'{o}w, Poland}\\
\mbox{$^{4}$Smoluchowski Institute of Physics, Jagiellonian University of Cracow, 30-059~Krak\'{o}w, Poland}\\
\mbox{$^{5}$GSI Helmholtzzentrum f\"{u}r Schwerionenforschung GmbH, 64291~Darmstadt, Germany}\\
\mbox{$^{6}$Technische Universit\"{a}t Darmstadt, 64289~Darmstadt, Germany}\\
\mbox{$^{7}$Institut f\"{u}r Strahlenphysik, Helmholtz-Zentrum Dresden-Rossendorf, 01314~Dresden, Germany}\\
\mbox{$^{8}$Institut f\"{u}r Kernphysik, Goethe-Universit\"{a}t, 60438 ~Frankfurt, Germany}\\
\mbox{$^{9}$Excellence Cluster 'Origin and Structure of the Universe' , 85748~Garching, Germany}\\
\mbox{$^{10}$Physik Department E62, Technische Universit\"{a}t M\"{u}nchen, 85748~Garching, Germany}\\
\mbox{$^{11}$II.Physikalisches Institut, Justus Liebig Universit\"{a}t Giessen, 35392~Giessen, Germany}\\
\mbox{$^{12}$Frederick University, 1036~Nicosia, Cyprus}\\
\mbox{$^{13}$Department of Physics, University of Cyprus, 1678~Nicosia, Cyprus}\\
\mbox{$^{14}$Laboratoire de Physique des 2 infinis Irène Joliot-Curie, Université Paris-Saclay, CNRS-IN2P3. , F-91405~Orsay , France}\\
\mbox{$^{15}$Nuclear Physics Institute, The Czech Academy of Sciences, 25068~Rez, Czech Republic}\\
\mbox{$^{16}$LabCAF. F. F\'{\i}sica, Univ. de Santiago de Compostela, 15706~Santiago de Compostela, Spain}\\
\mbox{$^{17}$Warsaw University of Technology, 00-662~Warsaw, Poland}\\
\\
\mbox{$^{a}$ also at Coimbra Polytechnic - ISEC, ~Coimbra, Portugal}\\
\mbox{$^{b}$ also at Technische Universit\"{a}t Dresden, 01062~Dresden, Germany}\\
\mbox{$^{c}$ also at Charles University, Faculty of Mathematics and Physics, 12116~Prague, Czech Republic}\\
\mbox{$^{d}$ also at University of Wroc{\l}aw, 50-204 ~Wroc{\l}aw, Poland}\\
} 
%\date{\today}

\begin{abstract}
The double differential production cross sections, $d^2\sigma/d\Omega dE$,  
for hydrogen isotopes and charged pions
in the reaction of p + Nb at 3.5 GeV proton beam energy have been measured by the  
High Acceptance DiElectron Spectrometer (HADES). Thanks to the high acceptance 
of HADES at forward emission angles and usage of its magnetic field, 
the measured energy range of 
hydrogen isotopes could be significantly extended 
in comparison to the relatively scarce experimental data available 
in the literature. The data provide information about the development 
of the intranuclear cascade in the proton-nucleus collisions. 
They can as well be utilized to study the rate of energy/momentum dissipation 
in the nuclear systems and the mechanism of elementary  
and composite particle production in excited nuclear matter at normal density. 
Data of this type are important also for technological and medical applications.
Our results are compared to models developed to describe the processes relevant 
to nuclear spallation (INCL++) or oriented 
to probe either the elementary hadronic processes in nuclear matter 
or the behavior of compressed nuclear matter (GiBUU).
\end{abstract}

\maketitle


\section{\label{sec:introduction} Introduction}
 
Proton-nucleus collisions are an important tool for the
investigation of complex phenomena in strong interaction physics.  In
particular, reactions with protons at a beam energy of a few GeV allow 
to study spallation reactions, in which the target nucleus
disintegrates into many smaller fragments and reaction products. 
A thorough understanding of their underlying mechanisms is 
relevant, in particular as these reactions are excellent tools for
fundamental and applied science  
\cite{SINQ,ISIS,SNS,ESS,CSNS}.  
One example for an important 
application of spallation reactions is nuclear waste transmutation in
Accelerator Driven Systems (ADS)  
\cite{SNS,JSNS,ESS,CSNS,Vandeplassche,ENEA_ADS,HINDAS,OECD_NEA}.
They are also crucial for the understanding 
of the nuclear spallation contribution to nucleosynthesis 
\cite{VangioniFlam2000365, Ramaty_1997ApJ,
VangioniFlam2001583,Mashnik_2000astro.ph,
https://doi.org/10.1111/j.1945-5100.2009.tb00746.x},  
cosmic ray propagation in the Galaxy \cite{Amato:2017dbs}  
or extensive air showers generated 
by high-energetic cosmic rays \cite{Engel:2011zzb}. 
Proton-nucleus collisions also provide a valuable
laboratory for the investigation of light nuclei formation 
in excited nuclear systems 
\cite{Mrowczynski_1}. 
Finally, they serve as an
essential baseline measurement for the interpretation of heavy-ion 
collision data with respect to dense nuclear matter.

A characteristic property of spallation reactions is the abundant
emission of neutrons. Due to this fact, it was possible to build
efficient sources of neutrons with controlled flux and energy
distribution (so called spallation sources). It is therefore not
surprising that neutron production in spallation reactions was
intensively studied experimentally, and nuclear models were developed
to parametrize neutron angular distributions and energy spectra in
interactions of protons with thin and thick targets, cf. e.g. 
\cite{LET00A,LER02A}.
These models were, however, not able to reproduce satisfactorily
the emission of protons and light charged particles like tritium and
helium isotopes which may strongly influence the stability of the
neutron source. Thus, subsequent experimental 
\cite{LET02A} 
and theoretical
\cite{BOU02A,BOU04A} 
investigations were undertaken for these purposes.
Furthermore, the emission of other products in spallation reactions,
like pions 
\cite{CAT08A},
as well as complex heavy nuclei, were studied both experimentally 
\cite{AUD06A,NAP07A,GIO13A} 
and theoretically 
\cite{AOU06A,BOU13A,SHA17A,SIN18A}. 
In spite of these
efforts, there remain problems which are not satisfactorily
explained by existing models 
\cite{BUD10A,FID17A}. 
It is therefore necessary to
continue the studies of spallation reactions to gain 
additional insight in the 
detailed mechanism of these processes.

According to Serber 
\cite{PhysRev.72.1114},  
proton-nucleus collisions proceed
in two steps. During the first, the dynamical one, the projectile
particle transfers its energy to the nuclear target in 
a cascade of binary collisions with the target constituents. During this stage, 
the production and emission of energetic particles is expected. The second
step consists of a statistical emission of slow particles from the
thermalized remnant of the target nucleus. Such a two-step picture of
the reaction agrees well with the angular
distributions of observed reaction products, which are found to be
isotropic for low-energy products and forward-peaked for high energy
ones. 
It also agrees with the properties of their energy spectra, which are of
Maxwellian shape for small and of exponential shape for large particle
energies. However, present day models are not able to
quantitatively reproduce the differential 
\cite{SHA16A}, 
as well as the total 
\cite{SHA17A,SIN18A} 
production cross sections of complex particles emerging from
spallation reactions. Such effects were observed for all target nuclei
starting from light ones, as e.g. Ni 
\cite{BUD09A}, 
through intermediate masses like Ag 
\cite{FID17A}, 
to heavy ones such as Au 
\cite{BUB07A} 
nuclei. Therefore, it seems indispensable to re-examine 
in more detail the emission mechanism
of the main products of spallation reactions, like nucleons and
pions. While the production of neutrons was investigated in great
detail, because of its technological applications, the proton and pion
data are not abundant. Especially data on proton and pion production cross 
sections measured simultaneously in one experiment are rare.
In the last years, such reactions were  
studied by the HARP 
(\cite{HARP_PS214_1}, references therein) and the HARP-CDP 
(\cite{HARP_CDP_Al_2012}, and references therein) collaborations.
 
The present investigation yields both the double differential 
cross sections for charged pions as well as for hydrogen isotopes produced
in collisions of a 3.5 GeV proton beam with a Nb target, measured
by the HADES collaboration 
\cite{HADES,HADES_Agakishiev_1}. 
The experimental data are confronted with the results 
of two commonly used models (INCL++ 
\cite{PhysRevC.87.014606} 
and GiBUU 
\cite{Buss_2011mx}). 
Our paper is organized as follows. In section II the experimental
setup of HADES is described and the parameters of most relevant parts
of the detection system are given. Section III presents the analysis
procedures applied in order to derive the double differential
cross sections of interest. Various
components of experimental uncertainty are discussed as well. In
section IV the full set of double differential cross sections 
for p, d, t, $\pi^{+}$ and $\pi^{-}$ is given. 
In section V, the verification of the analysis scheme applied in this
study is done by a comparison of the current results with examples
of similar data available in the literature. Section VI provides 
the main assumptions of the applied models. Methods of
evaluation of the cross sections are presented as well. 
The comparison of theoretical and experimental results 
for the currently examined reactions is discussed in section VII. 
The conclusions about validity of the models in their description 
of the studied collision dynamics and their predictive 
power are also given.  The summary of the work is presented 
in section VIII.

\section{\label{Setup} Experimental Setup}

\begin{figure}[!h]
\includegraphics[width=0.43\textwidth]{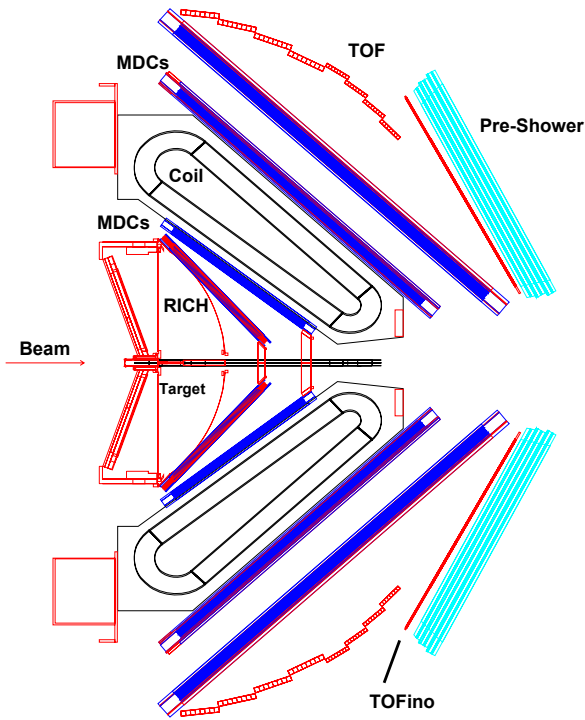}%
\caption{\label{HADES_setup2} The cross section of the HADES setup used during 
the measurement of p+Nb reactions at 3.5 GeV. The location of the most important 
detectors for the measurement of charged pions and hydrogen isotopes is shown 
relative to the beam axis. The particle identification was based on 
the d$E$/d$x$ vs. momentum measurements with the 
Multiwire Drift Chambers (MDC) 
and in the TOF/Tofino scintillating walls. The latter were used also as triggering detectors. 
For more details see the text.}
\end{figure}

The High Acceptance Dielectron Spectrometer (HADES)  
\cite{HADES,HADES_Agakishiev_1} 
of the Heavy-Ion Research Laboratory 
(GSI Helmholtz Center f\"ur Schwerionenforschung) Darmstadt, Germany, 
is optimized to perform the research with proton, 
pion and heavy-ion beams impinging on stationary solid or liquid targets. 
It provides information about production rates, angular and energy distributions 
for dileptons, mesons and baryonic products.
The presented results are derived from experimental data where a $^{93}$Nb target has been 
bombarded by 3.5 GeV energy protons 
\cite{AGAKISHIEV2012304,PhysRevC.88.024904,PhysRevC.90.054906,HADES_2014_Lambda_p_Nb,PhysRevLett.114.212301,PhysRevC.94.025201,HADES_2015oef,2018735}.
For the detection and identification of pions and hydrogen isotopes 
the most important components of the detection system are: target, 
Multiwire Drift Chambers (MDC) 
and scintillating walls called TOF and Tofino. 
The mutual positions of these detectors are shown in the cross section 
of the HADES setup presented in fig. \ref{HADES_setup2}.

A segmented solid target of $^{93}$Nb has been used in the present experiment. 
Its diameter was 2.5 mm and its total thickness was 0.45 mm. 

The detection system is organized in six identical sectors covering the complete 
azimuthal angle, except for the magnet coils and providing an acceptance 
between 18$^{\circ}$ and 85$^{\circ}$ in polar angle. 

The tracking system consists of 24 individual Multiwire Drift Chambers (MDC). 
They were filled with argon gas doped by isobutane as a quencher. 
The MDCs were operated at atmospheric pressure. 
Their position resolution is $\le$ 100 $\mu$m in polar direction and $\le$ 200 $\mu$m 
in azimuthal direction. 
In each detection sector there are two chambers in front and two chambers behind the magnet. 

The superconducting toroidal magnet provides a maximal magnetic field of 3.6 T and 
causes the momentum dependent bending of the trajectories of charged 
reaction products. 
The track reconstruction procedures permit a momentum resolution of $\delta{p}/p$ $=$ 4\%.  

Despite of the small effective thickness of the MDC system (0.5\% of a radiation length) 
it is possible to measure the energy loss of charged particles in the detector medium. 
It is done by means 
of the measured Time-over-Threshold \cite{Nygren,Kipnis} of a given signal. 
Taking into account the particle's path length the resulting d$E$/d$x$ 
resolution is better than 7\%.
This allows for the particle identification by means of the specific energy loss.

The TOF and Tofino detectors located at the end of the detection system were intended 
as stop detectors for the particle Time-of-Flight measurement. 
However, due to lacking start detectors during the measurement,  
they are used as triggering detectors and additional d$E$ detectors only.

The TOF scintillating wall covers polar angles from 44$^{\circ}$ to 85$^{\circ}$.  
The intrinsic time resolution of the scintillating strip is 150 ps 
and its position resolution 3 cm. 
The d$E$/d$x$ resolution for these scintillators is 4\%.

The Tofino covers the polar angles between 18$^{\circ}$ and 45$^{\circ}$. 
Its timing resolution is 420 ps. 
Since signals of these detectors are read out only 
at one side of the scintillating strips, 
the resulting energy loss resolution is 8\%. 
Tofino has also a worse double hit resolution than TOF.

The analysis of the detected data is accompanied by careful HGeant 
and HYDRA (Hades sYstem for Data Reduction and Analysis) 
\cite{Sanchez_2003, HADES_Agakishiev_1} simulations of the response 
of each part of the detection system including its acceptance, efficiency, 
tracking, energy loss and calibration.   

The detection system of HADES is described in more details in  
\cite{HADES_Agakishiev_1}.

\section{Particle selection and identification}

\subsection{\label{PID} Particle identification and background subtraction}

The particle identification (PID) and the background subtraction 
utilize the good energy loss resolutions 
of the MDCs and the TOF/Tofino scintillating walls.
A procedure of consecutive cut definitions, ranging from level 1 ($mass-momentum$ distribution) through 
level 2 ($dE/dx_{MDC}$ vs. $momentum$) to level 3 ($dE/dx_{TOF}$ vs. $momentum$ 
and $dE/dx_{Tofino}$ vs. $momentum$), has been developed. 

The $mass$ cut of level 1 provides just a rough separation of the $mass$ ranges 
of individual species ($\pi^{+}$, p, d, t) in a $mass$ vs. $momentum$ plot. 
Due to the lack of particle velocity measurements during the p + Nb data taking,  
the time of flight (T-o-F) of the selected particle needed for $mass$ calculation 
was reconstructed by the comparison
to the T-o-F of the fastest identified particle of the event.
But, as long as the single distributions of reaction products are of interest, 
the coincidences with other particles have to be disregarded.
For this reason the identification based on the $mass-momentum$
dependence for individual particles cannot be utilized 
for their exact identification in the current analysis. 
At levels 2 and 3, for each selected bin of laboratory emission 
angle of 3$^{\circ}$ width and momentum of 25 MeV/$c$ width, 
asymmetric Gaussian functions (eq. \ref{asym_gauss}), 
allowing for different widths for low- and high energy losses, 
are fitted to the $dE/dx$ distributions.
\begin{equation}
f_g(\frac{dE}{dx})=\begin{cases}
\frac{1}{\sigma_l\sqrt{2\pi}}e^{-\frac{1}{2}\left(\frac{\frac{dE}{dx}-\mu}{\sigma_l}\right)^2} \textnormal{for} &\frac{dE}{dx}-\mu\le0,\\
\frac{1}{\sigma_r\sqrt{2\pi}}e^{-\frac{1}{2}\left(\frac{\frac{dE}{dx}-\mu}{\sigma_r}\right)^2} \textnormal{for} &\frac{dE}{dx}-\mu>0.
\end{cases}
\label{asym_gauss}
\end{equation}
The width of the PID cut has been selected according to the value 
of the total standard deviation of the fitted asymmetric Gaussians:
%$\sigma = 0.5 \cdot ({\sigma_l} + {\sigma_r})$, 
$\sigma = \frac{({\sigma_l} + {\sigma_r})}{2}$, 
around the mean value $\mu$ of the fitted distribution.

In this way, the 2D cuts ($dE/dx - momentum$) 
for all positively charged reaction products of interest 
have been created and applied to the raw data in order to select 
the experimental distributions. The separation of negatively charged pions 
from other reaction products is provided by their opposite deflections in the 
magnetic field. There is no need for an additional particle identification, 
as contamination of $\pi^{-}$ with $K^{-}$ or $e^{-}$ 
is insignificant and disregarded here.

\begin{figure*}[ht]

\includegraphics[width=0.96\textwidth]{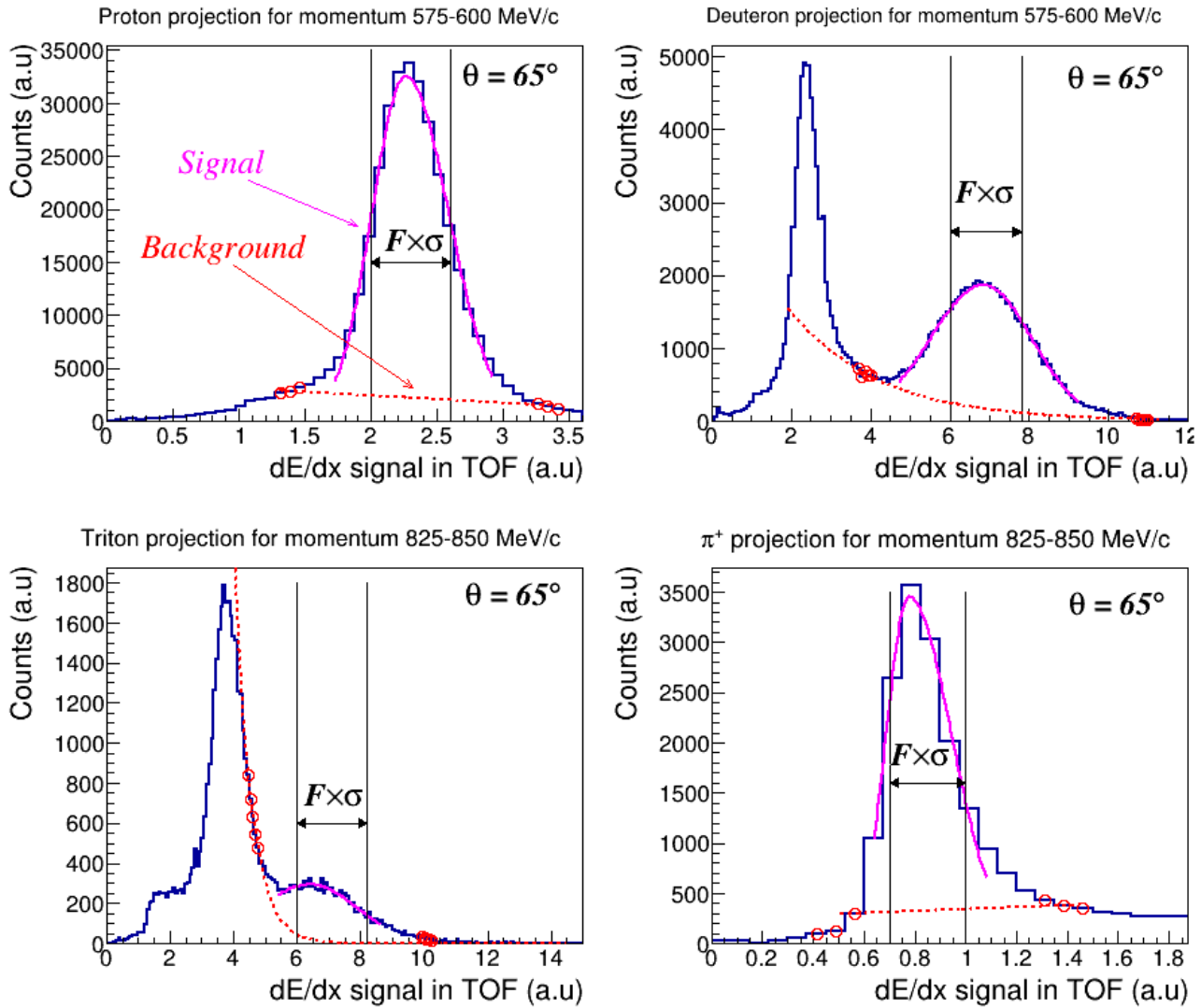}%
\caption{\label{PID_cut_1} Examples of the identification of protons (top-left), 
deuterons (top-right), tritons (down-left) and positively charged pions (down-right) 
for the emission angle $\theta = $ 65$^{\circ}$ and the momentum bin of 575-600 MeV/$c$. 
The functions fitted to the signal peaks are an asymmetrical Gaussian function 
according to Eq. (\ref{asym_gauss}), 
(continuous purple curves) whereas the underlying background 
is approximated with a straight line (for protons and $\pi^{+}$) 
or an exponential function (for deuterons and tritons) - dashed red curves. 
To the left of the deuteron and triton peaks the contamination 
by protons or deuterons, respectively, is visible.
The PID cuts of various widths equal to 
$F$ $\cdot$ $\sigma$ are applied 
to the signal component. $\sigma$ is the width of the respective fitted 
Gaussian distribution whereas the parameter $F$ takes values 
between 0.6 and 1.8 (see \ref{Uncertainties}). 
}

\end{figure*}

The distributions of selected particles,  
after their projection onto the $dE/dx_{TOF/Tofino}$ axis, 
are fitted with signal and background distributions. 
It is done again for each momentum bin of 25 MeV/$c$.
The background component originating from misidentified 
"neighboring" particles is subtracted. 
Examples of identification of protons, deuterons, 
tritons and positively charged pions  
at the emission angle of 65$^{\circ}$ and for one momentum bin 
of 575-600 MeV/$c$ are presented in fig. \ref{PID_cut_1}.

The ranges of particle momenta considered for the calculation of cross-section
have been restricted by the demand that the resulting background-to-signal
ratio is not larger than 0.1 - for protons
and positively charged pions, and 0.6 - for deuterons and tritons, 
extracted in a $\pm$ 1 $\sigma$ range of the Gaussian function 
fitted to the signal peak, respectively. 

In the HADES apparatus also secondary particles 
emerging from non-target material are created. 
Their contribution to the spectra of primary reaction products 
is suppressed by the particle selection in the tracking procedure 
and can therefore be ignored for 
cross sections of single particles.

\subsection{\label{Eff_corr} Determination of efficiency}

The overall efficiency has to be taken into account 
in the absolute normalization of the obtained distributions. 
Here we define it as the combination 
of the geometrical detector acceptance (Acc) and efficiency (Eff),  
where efficiency includes the   
track reconstruction efficiency, PID efficiency, trigger conditions 
and data acquisition efficiency.

The overall efficiency is calculated using standard simulation 
tools of HADES - HGeant + HYDRA \cite{Sanchez_2003,HADES_Agakishiev_1}. 
For generating the initial distributions 
of charged products in p (3.5 GeV) + Nb reactions, the INCL++ model 
has been applied. It provides realistic yields 
and distributions of the dominant reaction products. 

With the use of the selected event generator the simulated distributions 
of emission angle $\theta$ vs. $momentum$ for individual particles are created: 
the so-called "initial" ones (without taking into account the HADES apparatus) 
and the "real" ones (with the inclusion 
of the full response of HADES). 
The predefined particle identification cuts are applied 
for the "real" $\theta$ vs. $momentum$ distributions. 
By dividing the "real" distribution by the "initial" one, 
the overall efficiency is calculated, bin-by-bin,  
for each reaction product of interest.
The overall efficiency of the HADES system is angle- and energy dependent. 
Thus, the calculated factors 
are applied to the values of absolute cross section for each  emission 
angle and for each individual energy bin.

\begin{figure}[!h]
\includegraphics[width=0.5\textwidth] {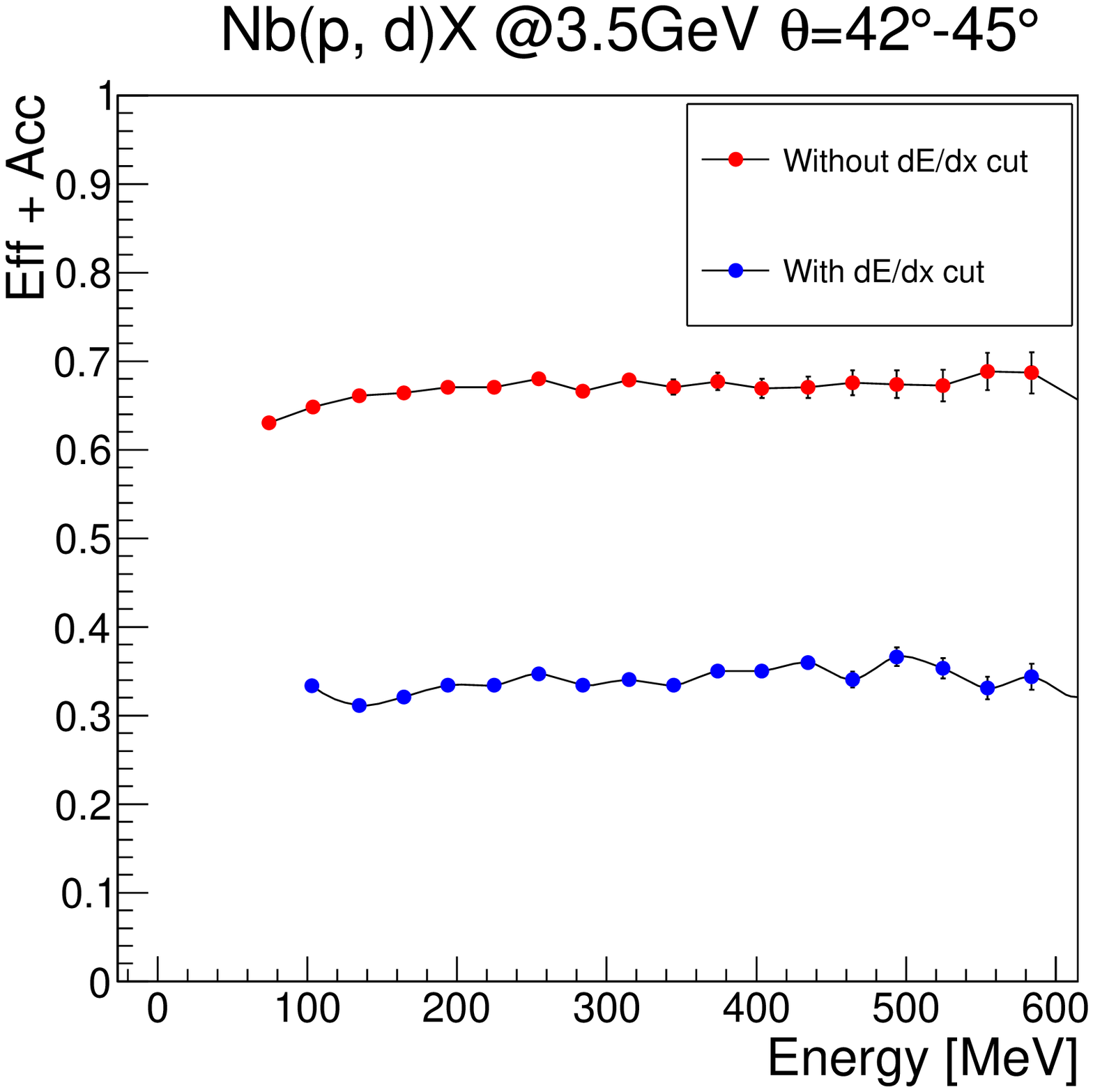}%
\caption{\label{effi_d_45deg} 
Example for the combined efficiency and acceptance 
of deuterons at 42$^{\circ}$ $<$ $\theta$ $<$ 45$^{\circ}$ 
laboratory angle. 
The red dots represent the efficiency obtained without PID cuts, 
whereas the blue dots show the final efficiency. 
In this case, the PID cuts were applied to the 
"real" distributions of deuterons. For details see the text.
}
\end{figure}

Figure \ref{effi_d_45deg} shows an example of calculated overall efficiency depending 
on the the energy of detected deuterons for the angular bin 
of 42$^{\circ}$ $<$ $\theta$ $<$ 45$^{\circ}$, before 
and after application of the PID cut to the "real" distribution. 

In the present studies, the possible modification of overall 
efficiency for purely inclusive spectra due to contribution 
of secondary particles is solved by tracking procedures.
The secondaries registered in the triggering detectors create a trigger bias.
For the calculation of the overall efficiency, the secondary particles from events
generated by INCL++ are "created" by HGeant and effectively tracked by
HYDRA back to their vertices. Those not originating from 
the target are suppressed. 
This is sufficient for a reliable determination of efficiency 
by dividing the "real" by "initial" distribution. 

The possible bias due to the trigger condition,
requiring at least three charged particles registered 
in TOF/Tofino walls, on the overall efficiency 
for single spectra is taken into account in the simulations 
of the "real" distributions 
with the use of the event generator, HGeant and HYDRA.  
Thus, the so-called trigger bias on the single spectra 
does not need to be treated separately.

\subsection{\label{Crosssec} Calculation of cross section}

The recorded multiplicity has been calculated for each particle of interest and for each  
selected energy- and angular bin by subtracting the background 
from the signal distribution and by integrating the differences. 
The obtained numbers were corrected by the calculated overall efficiency 
(efficiency$\times$acceptance) values.
For the calculation of the absolute value of the cross sections the normalisation 
factor derived in the former analysis of HADES 
$\pi^{-}$ data has been used \cite{M_Lorenz_thesis,PhysRevC.88.024904,Tlusty}. 
This normalization factor was obtained by interpolating 
known pion production cross sections 
\cite{HARP_CDP_Be2_2009,HARP_CDP_Ta_2009,HARP_CDP_Cu_2009,HARP_CDP_Pb_2010}. 
As a result of this interpolation we determine a total reaction cross section of 
$\sigma_{pNb}$ = 848$\pm$127 mb. A detailed description of the
procedure is given in \cite{Tlusty}.

\subsection{\label{Uncertainties} Uncertainties}

The statistics of the collected data is very high. The total number 
of analyzed events amounts to about 10$^{8}$.
For each energy bin of the presented data the statistical error 
is negligible and is therefore not shown in the plots.

The contributions to the systematic uncertainties are: 
\begin{itemize}
\item uncertainty of particle identification; 
\item uncertainty of overall efficiency;
\item differences of the response of individual sectors 
of the HADES detection system;
\item uncertainty of the absolute normalization factor.  
\end{itemize}

Only the last component of the systematic error 
is independent of phase space and amounts to 15\%. 
This was established in former analyses of HADES pion spectra 
from p (3.5 GeV) + Nb \cite{PhysRevC.88.024904} and their comparison to similar 
results by the HARP-CDP collaboration \cite{Tlusty}. Other components are energy 
and emission angle dependent, and will be shortly discussed below.  

\subsubsection{\label{pid/back} Particle identification}

Due to the lacking $mass$ identification based on the T-o-F information,  
the deconvolution of the signal and background tracks is based 
only on the specific energy loss method.
The resolving power of this approach is limited 
and varies with the energy of particles searched for.
In order to study the level of signal/background misidentification,  
various widths of identification cuts have been used. 
The applied cut widths were calculated by multiplication of the standard deviation 
$\sigma$ of the fitted asymmetrical Gaussian function by a factor $F$ equal to  
0.6, 0.8, 1.0, 1.2, 1.5, 1.8 (cf. fig \ref{PID_cut_1}). 
For each of the applied cut widths the quantity of the signal counts 
is calculated by the subtraction of the background contribution from the signal one. 
The standard deviation of the average of all obtained results 
for the given particle, emission angle and energy bin  
is assigned as the systematic uncertainty of the PID procedure. 
For protons this component of the systematic uncertainty 
does not exceed 5\% for almost the complete energy range. 
The largest values of 12\% appear for the highest energies 
for tritons.

\subsubsection{\label{eff} Efficiency}

Since the method of overall efficiency calculations consists 
in dividing two simulated distributions, the effects of minor imperfections 
of the used model cancel out. 
In the current studies only the energy regions where the overall efficiency changes 
monotonically are selected for the further analysis. 
The small fluctuations observed within the selected energy limits 
of the overall efficiency have been smoothed by applying a sliding average 
of three consecutive energy bins. The standard deviation of the sliding average 
is assigned as the systematic error of the overall efficiency. 
Its value varies in the range 2 - 5\%.

\subsubsection{\label{sect} Sector}

As explained in section \ref{Setup}, the HADES detection system consists 
of six equivalent sectors, which cover the forward emission cone and provide 
detection acceptance over the full azimuth angle $\phi$. 
It was checked whether all sectors give equivalent contribution to the measured cross sections.
For this purpose, the same kind of analysis as described above for the global setup has been performed 
for the particles detected in each individual sector. 
The differences are again dependent on the kind of detected particle, its energy and the emission angle.
The standard deviation of the average of results for individual sectors has been calculated 
for the selected particle, emission angle and the energy bin. 
This value estimates the systematic uncertainty
resulting from the differences in performance of individual sectors
and is found to be below 7\%. \\

The total systematic error squared is calculated as 
the quadratic sum of the uncertainty components discussed above.
It is done for each particle type, selected emission 
angle $\theta$ $\pm$ 1.5$^{\circ}$ and 
for each energy bin (25 MeV) of the distribution.

\section{\label{Experimental_results} Experimental results}

The high event statistics collected during the p (3.5 GeV) + $^{93}$Nb run 
permits the determination of $d^2\sigma/d\Omega dE$ distributions for
p,  d,  t, $\pi^{+}$ and $\pi^{-}$ in the polar angular range from
20$^{\circ}$ to 80$^{\circ}$. 

\begin{figure}[!hbt]
\includegraphics[width=0.5\textwidth] {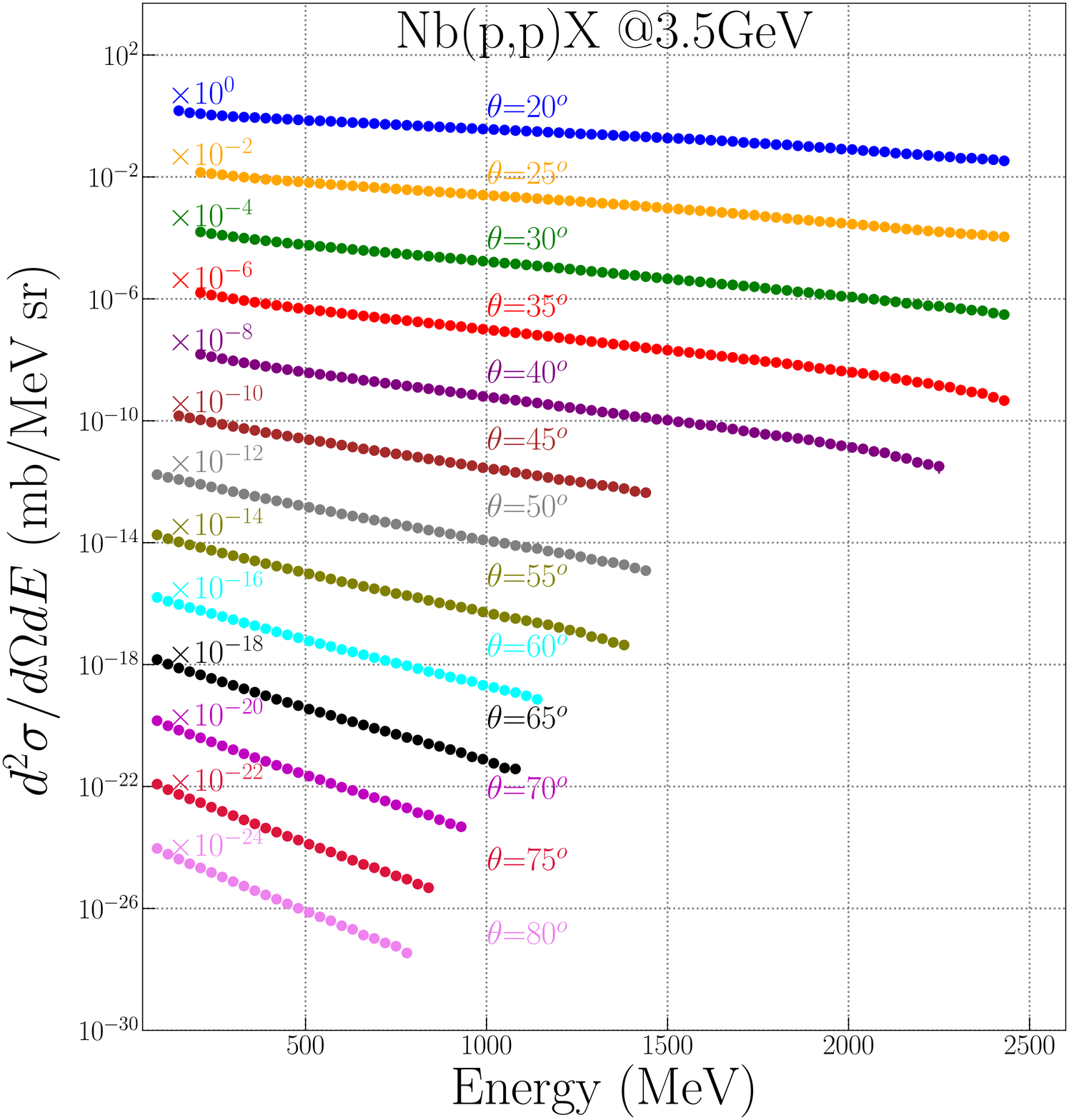} 
\caption{\label{p_all}
Double differential production cross sections of protons measured  
with HADES 
in p (3.5 GeV) + $^{93}$Nb reactions (full circles).
The distributions are scaled for better visibility.
Statistical errors are negligible; for systematic errors see text.
}
\end{figure}

\begin{figure}[!hbt]
\includegraphics[width=0.5\textwidth] {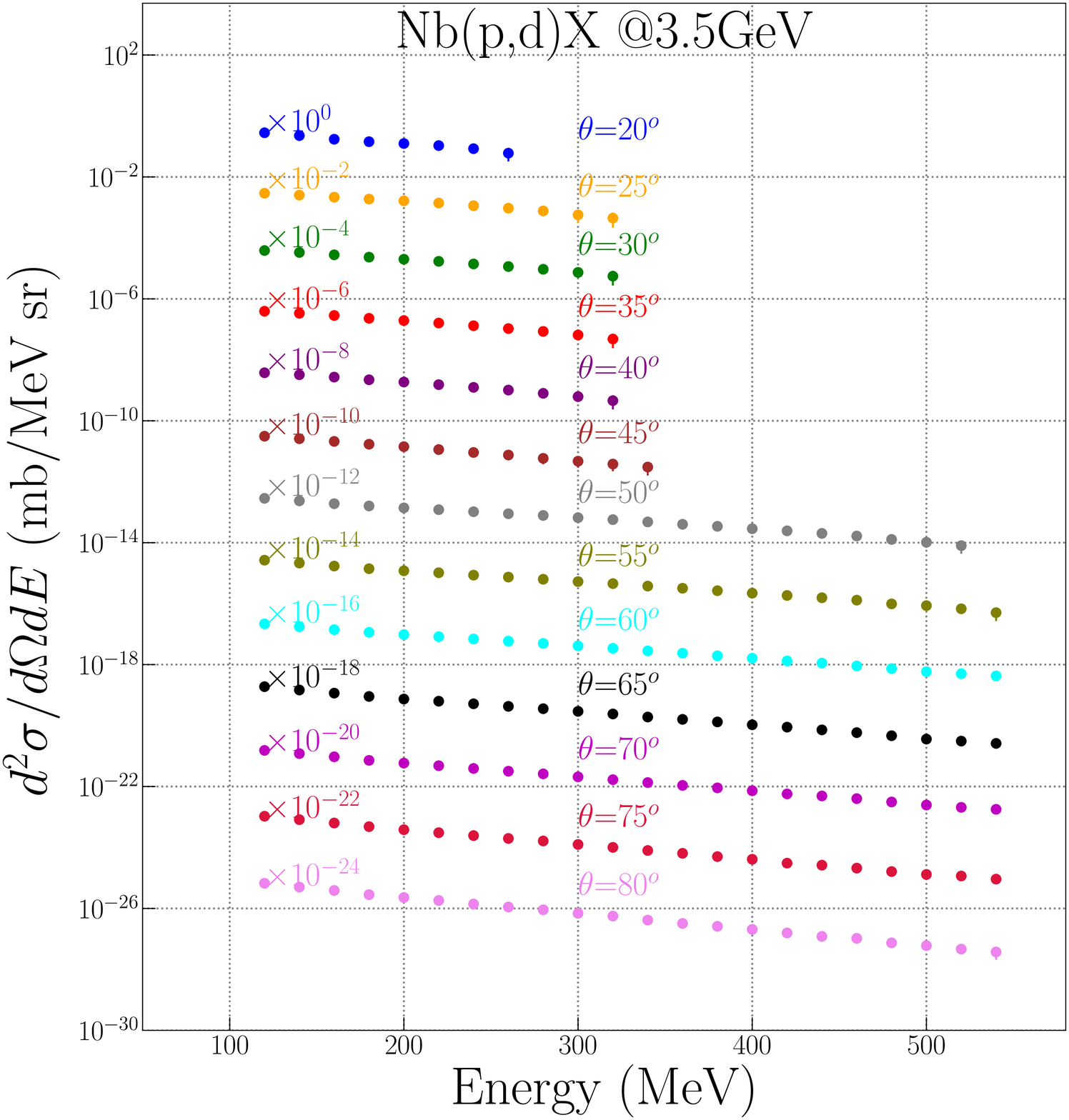} 
\caption{\label{d_all}
The same as in fig. \ref{p_all} but for deuterons.
}
\end{figure}

\begin{figure}[!hbt]
\includegraphics[width=0.5\textwidth] {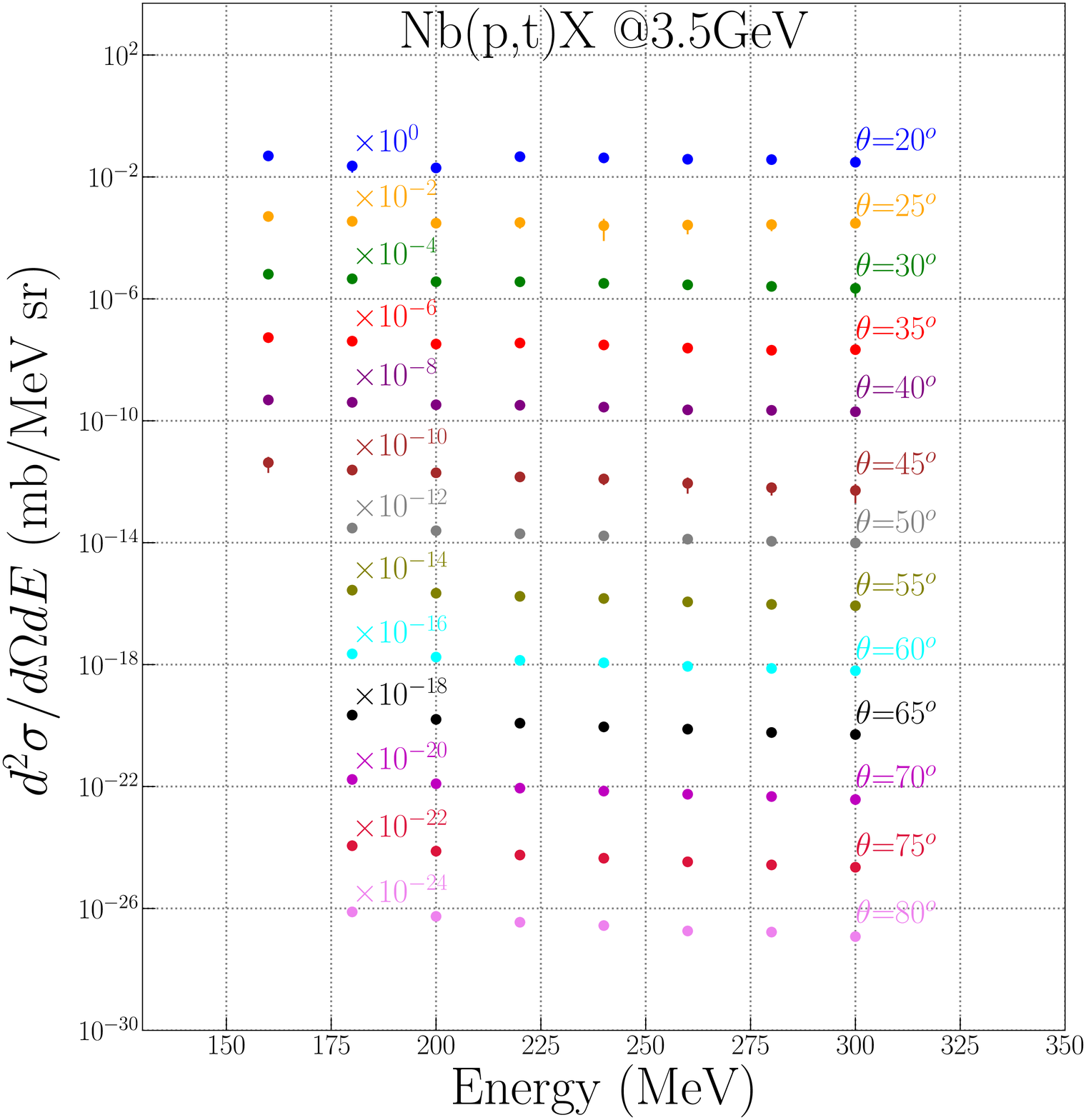} 
\caption{\label{t_all}
The same as in fig. \ref{p_all} but for tritons.
}
\end{figure}

\begin{figure}[!hbt]
\includegraphics[width=0.5\textwidth] {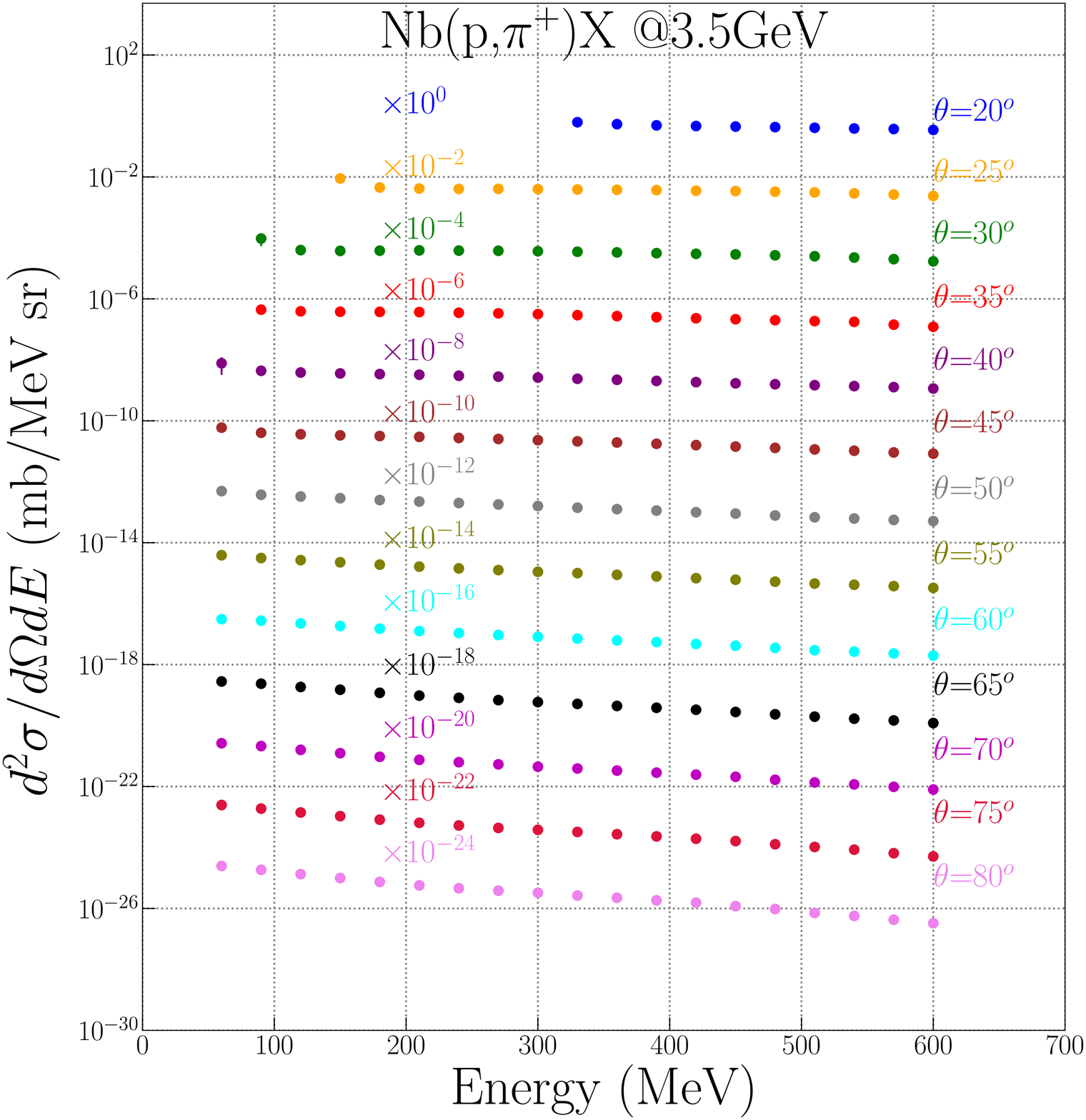}
\caption{\label{pi_plus_all}
The same as in fig. \ref{p_all} but for $\pi^{+}$. 
}
\end{figure}

\begin{figure}[!hbt]
\includegraphics[width=0.5\textwidth] {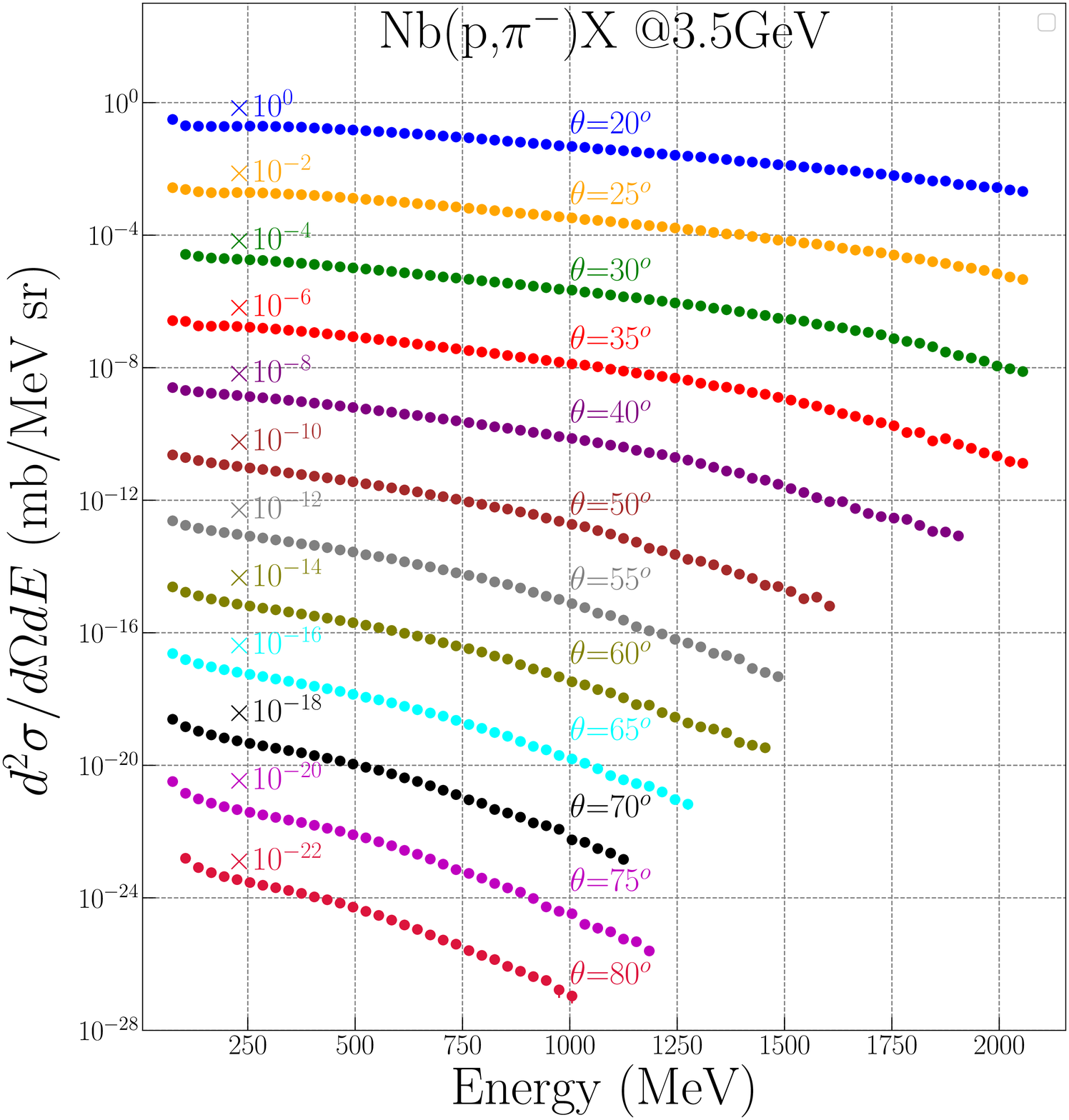}
\caption{\label{pi_minus_all}
The same as in fig. \ref{p_all} but for $\pi^{-}$. 
}
\end{figure}

All distributions for hydrogen isotopes are presented in figs. 
\ref{p_all}, \ref{d_all}, \ref{t_all},
whereas the cross sections for charged pions are shown in figs. \ref{pi_plus_all} and \ref{pi_minus_all}.
The experimental cross sections are plotted for the mean emission angles $\theta$  
of 20$^{\circ}$ - 80$^{\circ}$ in steps of 5$^{\circ}$ 
with spread of $\theta$ = $\pm$ 1.5$^{\circ}$. 
The presented values of cross sections are obtained for kinetic 
energy bins of 25 MeV size. 
The values of experimental errors are of the size of the marker 
and thus usually not visible in the plots. 
The constant normalization uncertainty of 15\% is not included 
in the error bars shown in the plots.
The insignificant statistical errors are neglected.
Differences in the energy range for specific reaction products 
at different emission angles result from 
geometrical acceptance limits.

Exploiting the magnetic spectrometer component of HADES it was possible 
to register and identify the charged particles over a much broader 
energy range than accessible in earlier experiments designed for
measurements of production cross section of light nuclear products. 
In fact, the energy range of the cross sections presented here in fig. \ref{p_all} 
clearly exceeds the energy limits of all proton distributions previously
available in the literature.


The experimental double differential cross sections of deuterons 
are limited in energy due the overlap of their
$dE/dx$ distributions with the ones of other hydrogen isotopes at higher particles
energy. Still, our results (fig. \ref{d_all}) 
on the cross section cover a wider range of energy 
than for deuteron data available up to now. The same holds for cross
sections of tritons (fig. \ref{t_all}).


The numerical values of the cross sections are available in the HEPData repository \cite{hepdata}.

\section{\label{verification} Comparison to world data} 

From the particles of interest of the current study only the negatively charged pions 
were examined in former analyses of HADES data collected for p+Nb reaction 
at 3.5 GeV \cite{PhysRevC.88.024904}. 
It was checked that the current analysis provides almost identical distribution  
for the transverse momenta, $p_{\bot}$, and the transverse mass, $m_{\bot}$,
as those published in \cite{PhysRevC.88.024904}. 

Since other data for p (3.5 GeV) + Nb reactions are not available in the literature, 
the data closest to the proton beam energy and the target mass have been selected 
for the verification of the current results. The shapes of the spallation spectra 
in the energy and mass range of interest are independent of the 
target mass and the proton beam energy. 
The magnitude of the cross section rises with both 
the beam energy and the mass number A 
of the target. Since this rise is weak it allows 
for a comparison of the results for similar 
target masses and beam energies. Usually, 
the experimental uncertainties of the compared distributions 
are of similar order as the expected differences of the cross sections.

\subsection{\label{spal_data} Low-energy spallation data}

The double differential cross sections for light charged nuclear products were measured 
in a few dedicated experiments. 
Here a comparison is preformed with the results obtained 
for proton-nucleus collisions by the PISA and HARP-CDP collaborations. 
Data provided by PISA cover a broad range of target nuclei 
(from C to Au) bombarded by protons of 1.2, 1.9 and 2.5 GeV energy 
\cite{BUB07A,BUD08A,BUD09A,BUD10A,FID14A,FID17A}. 
The HARP-CDP collaboration provided proton and pion spectra 
for proton induced reactions on some atomic nuclei from Be to Pb 
at 4.1 GeV proton bombarding energy \cite{HARP_CDP_Be1_2009,HARP_CDP_Be2_2009,HARP_CDP_Ta_2009,HARP_CDP_Cu_2009,HARP_CDP_Pb_2010,
HARP_CDP_C_2010,HARP_CDP_Sn_2011,HARP_CDP_Al_2012,Kang}.

\begin{figure}
\includegraphics[width=0.45\textwidth]{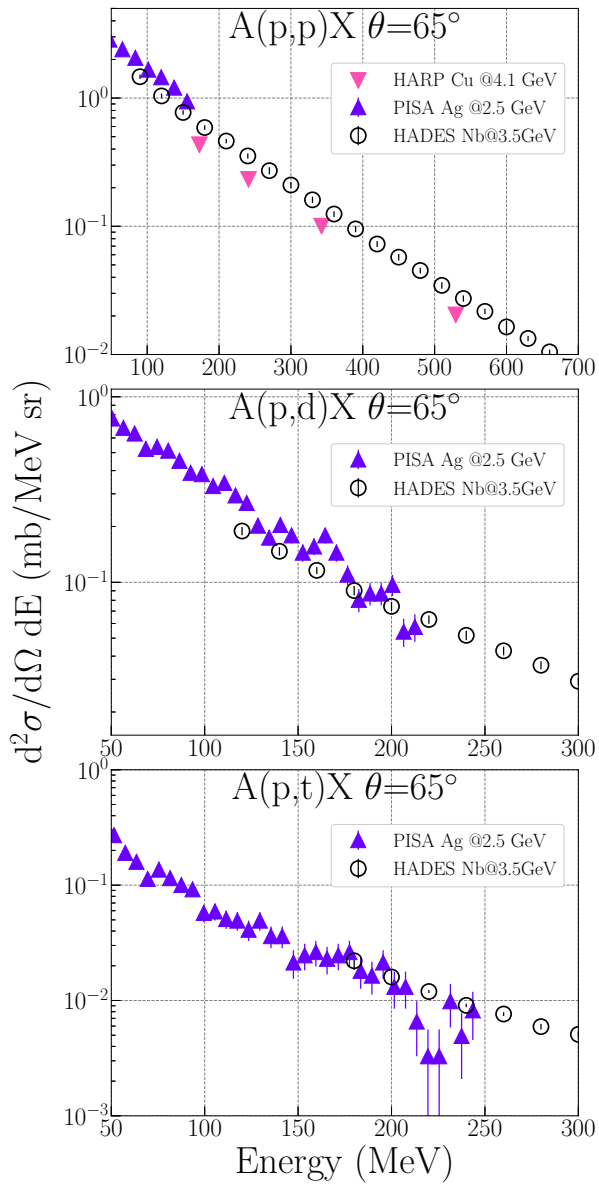}
\caption{\label{Comp_PISA_HARP_pdt} 
Examples of double differential cross sections for p (upper panel), 
d (middle panel) and t (lower panel) measured by HADES 
at $\theta$ = 65$^{\circ}$ laboratory emission angle 
in p (3.5 GeV) + $^{93}$Nb reactions.  
They are confronted with former results of the spallation  
experiment PISA \cite{FID17A} for the same isotopes and detection angle. 
The PISA data were measured with a $^{nat}$Ag target and 
for a proton beam energy of 2.5 GeV. The comparison 
results in a good agreement of the magnitudes 
and shapes of the distributions of both experiments. 
The double differential production 
cross sections for p are also compared with the results 
obtained by the HARP-CDP experiment 
for the same detection angle, but for the reaction   
p+$^{65}$Cu at 4.1 GeV \cite{HARP_CDP_Cu_2009}. 
The small differences in the magnitudes 
of the p distributions of the various experiments are due to the expected 
target mass and proton beam energy dependence of the production cross section. 
}
\end{figure}

In fig. \ref{Comp_PISA_HARP_pdt} the example of production cross sections for 
p (upper panel), d (middle panel) and t (lower panel) measured 
by the HADES  collaboration for p (3.5 GeV) + $^{93}$Nb 
and registered at a laboratory emission 
angle $\theta$ = 65$^{\circ}$ are presented. 
The HADES results for protons are compared to the results 
by PISA measured for the reaction 
of p + $^{nat}$Ag at 2.5 GeV \cite{FID17A} and the results of HARP-CDP 
registered for p + $^{64}$Cu reactions at 4.1 GeV proton 
energy \cite{HARP_CDP_Cu_2009}. 
Taking into account the differences in beam 
energies the results are in very good agreement.
HADES deuteron and triton production cross sections are compared 
to the results by PISA 
collected for p + $^{nat}$Ag at 2.5 GeV reactions \cite{FID17A}.
Here also a good agreement between HADES and PISA results is observed. 

\subsection{\label{pion_data} Mid-energy pion spectra}

\begin{figure}[!h]
\includegraphics[width=0.45\textwidth] {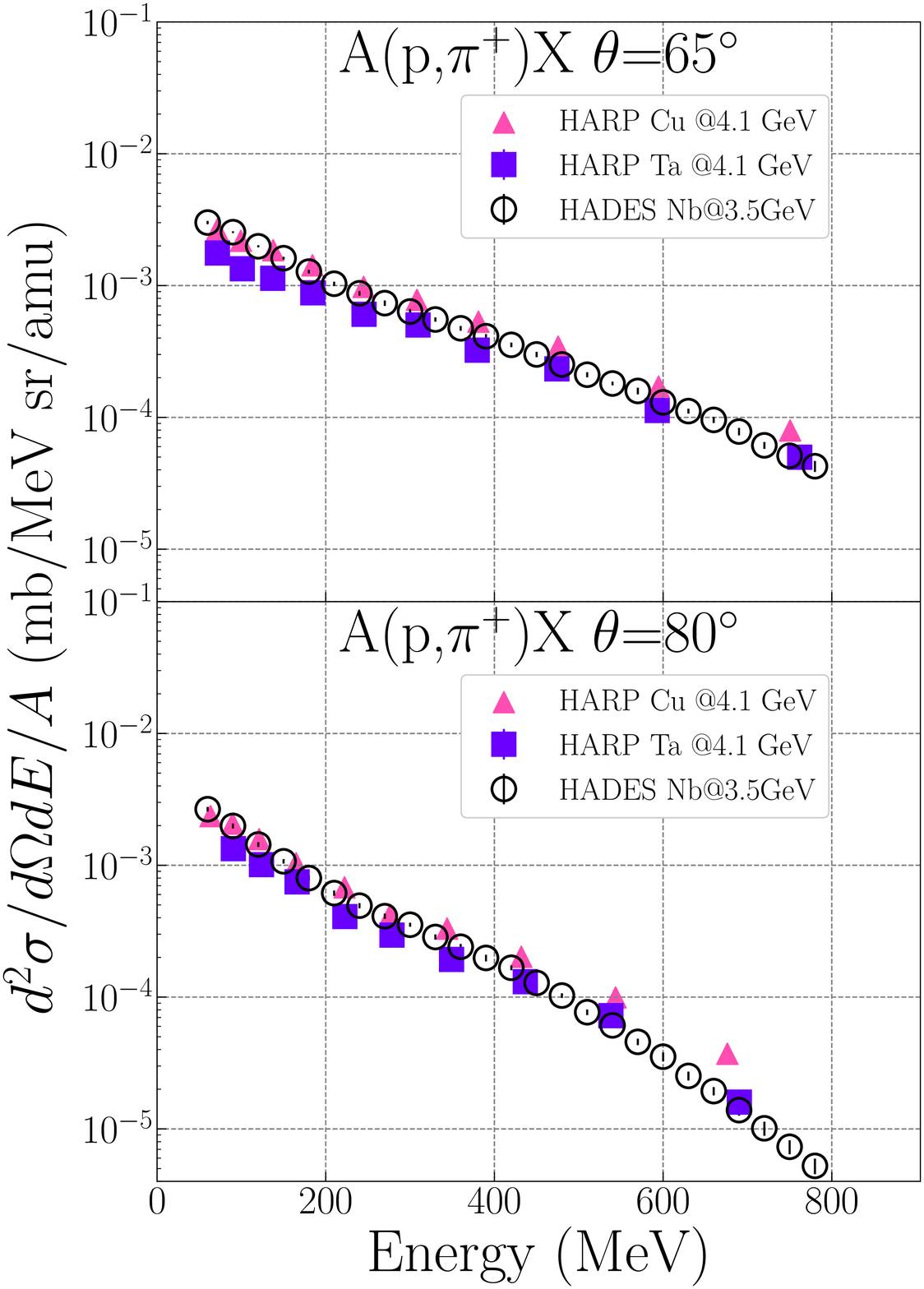}%
\caption{\label{Comp_HARP_pip} 
Examples of double differential cross sections measured at two emission angles 
($\theta$ = 65$^{\circ}$ - upper panel, and $\theta$ = 80$^{\circ}$ - lower panel) 
by HADES for $\pi^{+}$ for p (3.5 GeV) + $^{93}$Nb reactions.  
They are compared with similar results by the HARP-CDP 
collaboration measured for 4.1 GeV proton 
beam energy and targets of $^{64}$Cu 
\cite{HARP_CDP_Cu_2009} and $^{181}$Ta 
\cite{HARP_CDP_Ta_2009}.
}
\end{figure}

The comparison of $\pi^{+}$ production with results by the  
HARP-CDP experiment is shown in fig. \ref{Comp_HARP_pip}. 
The HARP-CDP data were collected for a proton beam energy of 4.1 GeV and targets of  
$^{64}$Cu \cite{HARP_CDP_Cu_2009} and $^{181}$Ta \cite{HARP_CDP_Ta_2009}. 
The shapes of the energy distributions of $\pi^{+}$ measured 
at two angles (65$^{\circ}$ and 80$^{\circ}$) 
are practically the same for all three targets.
Since the proton beam energies in both experiments were similar, 
the measured cross sections were additionally divided 
by the corresponding target mass numbers. 
It could be concluded that, in the examined range of target masses 
and proton energies, the deviation from expected cross section scaling 
with the target mass is lower than factor two. 
A good agreement of cross sections measured in both experiments is confirmed.

\section{\label{data_theory_comp} Comparison with models}

It is reasonable to expect that the angular distribution of particles,  
emitted in forward direction 
as well as their energy distributions, 
contain information of the first stage of the proton-nucleus 
collision.
This stage is referred to as an intranuclear cascade and is assumed 
to be a sequence of nucleon-nucleon and pion-nucleon 
interactions induced by the first collision of the projectile 
with one of the target constituents.

The obtained data provide a chance to extract important information
on this stage of the collision, especially as the main charged
participants of this process, pions and 
protons, are observed simultaneously in our experiment. 
They are accompanied with heavier hydrogen 
isotopes (d, t) with energies clearly higher 
than those typical for evaporated particles. 
Thus, most probably the observed deuterons and tritons 
originate as well from the first stage 
of the reaction.

The experimental distributions are compared to the
predictions of two models: GiBUU (release 2021, Feb 8, 2021) 
\cite{Buss_2011mx} and INCL++ (version v6.29-9198542) \cite{PhysRevC.87.014606},  
which are commonly used in investigations
of nucleus-nucleus collisions at GeV/A energies. The models 
differ in the level of approximation to the physical
phenomena appearing in the quantum-mechanical realm 
of dynamical nuclear systems.

The Intranuclear Cascade Model of  Li{\'e}ge (INCL++) 
\cite{PhysRevC.87.014606} (and references therein) has been
developed over the last four decades as a tool for simulations of
spallation reactions. It employs a semiclassical treatment of
the target nucleus and the nuclear cascade. An isospin and energy
dependent nuclear mean field is assumed in which particle
propagation proceeds along straight lines. Nucleon-nucleon as
well as pion-nucleon collisions are probed stochastically. Their
probability and the specific reaction channel depend 
on parameterized cross sections
known from the interaction of free hadrons. INCL++ introduces 
basic quantum-mechanical prerequisites like the Pauli blocking or
the tunneling probability of the Coulomb barrier by escaping particles
from the interaction region.

Despite of its relative simplicity this model has significant
advantages in comparison to other widely used models of such kind of
reactions: (i) Great attention is paid to the selection of
parameters of the target nucleus. It concerns the density profile,
diffuseness of the nuclear edge, the neutron skin and initial Fermi
momentum distribution of nucleons. (ii) During the cascade the
stability of the target nucleus is assured. The struck nucleus
undergoes a mass loss due to emission of particles but the heavy
remnant remains stable and does not blow up. (iii) For this reason, 
INCL++ allows to calculate the properties of the reaction remnant.
Its further fate can be credibly simulated by means of a 
statistical model. (iv) This model tries to explicitly introduce the
dynamic creation of composite nuclear products by means of the 
so-called surface coalescence \cite{LET02A,BOU04A,PhysRevC.87.014606}.

With the hypothesis of surface coalescence, the clustering is realized during 
the intranuclear cascade. The creation of composite particles is tested 
when the single nucleon is going to be emitted. Then other nucleons 
of suitable isospin are searched for in its phase space vicinity. 
Out of them the cluster is composed and,  
if the emission criteria for the new object are satisfied,  
it departs from the target nucleus.

In the GiBUU (Giessen Boltzmann-Uehling-Uhlenbeck project) model 
the multi-particle problem is discretized by the introduction of a
statistically significant set of test particles for each 
simulated real particle instead of probing the continuous 
probability distributions of nuclear systems. 
Carefully constructed Hamiltonian and 
collision terms are used in order to simulate the evolution of
the colliding system by solving transport equations. The time
dependent mean-field potentials - hadronic and electromagnetic - 
are included. As in other models, the geometrical cross sections 
(experimental or theoretical ones) for free hadrons are
used in the calculation of the collision probability and its type. 
Pauli blocking is included for the collision
output channel.

GiBUU is not equipped with mechanisms 
for the creation of composite particles. This is due to 
intrinsic limitations, namely the difficulty to create 
density fluctuations in the distributions of the individual test
particles. 

Default settings were assumed for the GiBUU 
and INCL++ models for all numerical calculations
performed in the present study.

\subsection{\label{p_and_pi} Protons and pions}

This subsection is devoted to angular and energy distributions of
the main charged reaction products, 
namely protons and charged pions. The light composite particles 
(deuterons and tritons), whose origin is much less 
understood, are considered afterwards. The cross sections presented 
here are given only for three detection angles, $\theta$ = 25$^{\circ}$,
55$^{\circ}$ and 80$^{\circ}$, because the angular dependence of the
data is monotonic and smooth. Furthermore, a selection of three
energy spectra was done in order to facilitate the observation of
certain trends of theoretical results which may differ for individual
models and for various detection angles.

The uncertainties indicated for all presented experimental data
include only the energy and angle dependent components of the systematic
uncertainties (see section \ref{Uncertainties}). The constant
component of the uncertainty of 15\%, resulting from absolute normalization
factor, is not included in the plots. The insignificant statistical
errors are neglected.

\begin{figure}[!h]
\includegraphics[width=0.5\textwidth] {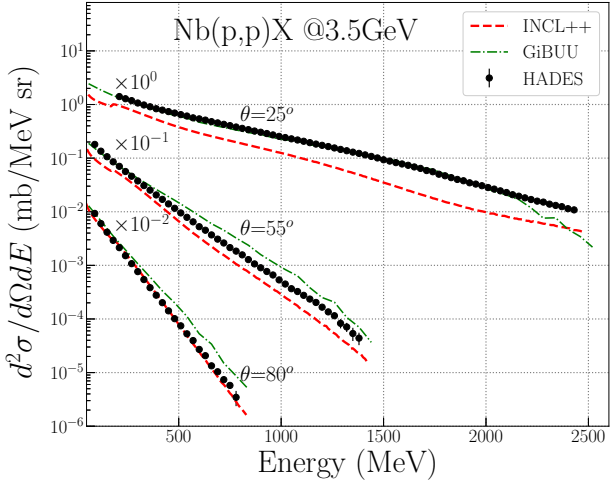}%
\caption{\label{p_m_3a} Double differential cross sections of protons 
measured by HADES in p (3.5 GeV) + $^{93}$Nb reactions (full circles). 
Cross sections are shown for three laboratory
emission angles $\theta$ = 25$^{\circ}$, $\theta$ = 55$^{\circ}$
(multiplied by factor 10$^{-1}$) and $\theta$ = 80$^{\circ}$
(multiplied by factor 10$^{-2}$). The experimental distributions are
compared to the results of two models: GiBUU (dash-dotted
lines) and INCL++ (dashed lines). The constant
normalization error of experimental data of 15\% is not shown.}
\end{figure}

\subsubsection{\label{prot} Protons}

The obtained distributions, which are shown in fig. \ref{p_m_3a},
vary monotonically over the whole investigated energy range. Their slopes
increase with the polar emission angle $\theta$.
A good agreement with the proton data at forward emission angles
is provided by the GiBUU model. The theoretical curve follows the data 
for $\theta = $ 25$^{\circ}$ in the whole presented energy range. 
The INCL++ model provides spectra of very similar shape as GiBUU, but 
underestimates the magnitude of the data by a factor larger than two.
The disagreement of the INCL++ model with the cross sections
measured for protons is smallest at the lowest available energies and
increases with increasing kinetic energy of the proton.
Thus, the model distributions are steeper than the
experimental curves. With increasing emission angle,  
GiBUU starts to overestimate the data, whereas the predictions of 
INCL++ get closer to the experimental distributions. For
the highest angle, $\theta$ = 80$^{\circ}$, a better
description of the data is provided by INCL++, whereas  
GiBUU overestimates the experimental cross section. The discrepancy
increases with the proton energy, attaining a factor $\approx$2 at the
edge of available data range.

When comparing the predictions of GiBUU on proton production 
it has to be taken into account that this model does not include  
the formation of composite nuclear particles, 
which will affect the yield and kinematic distributions of protons. 

\subsubsection{\label{charged_pi} Charged pions}

\begin{figure}[!h]
\includegraphics[width=0.5\textwidth] {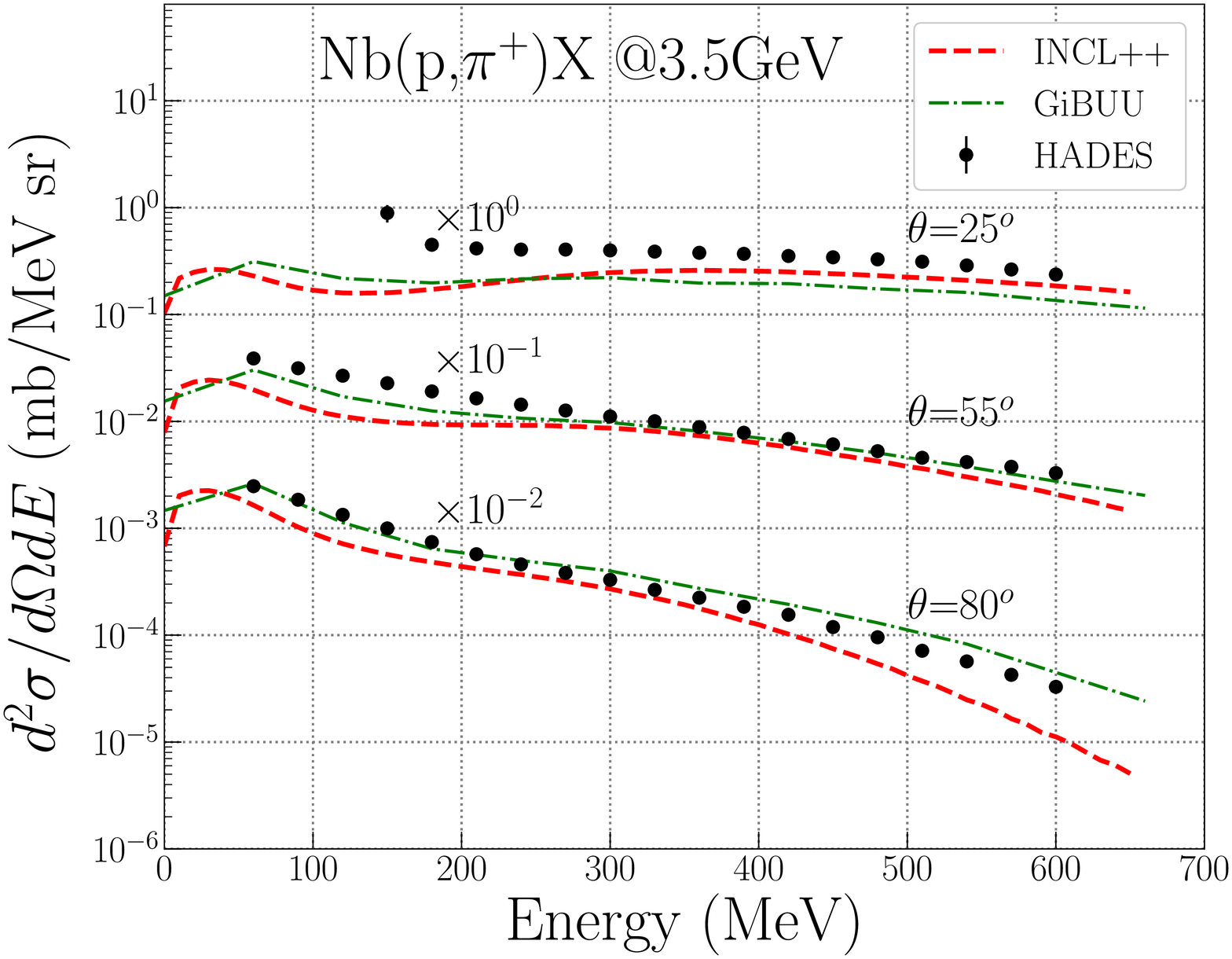}%
\caption{\label{pip_m_3a} The same as in fig. \ref{p_m_3a} but for
$\pi^{+}$.
}
\end{figure}

At the most forward emission angles of $\pi^{+}$ (see fig.
\ref{pip_m_3a}) the model distributions  underestimate the
experimental cross sections at least by a factor 2. Only for
higher particles energies above 500 MeV, results
of GiBUU and INCL++ follow the data.

The agreement improves for larger emission angles. Both models follow 
approximately the shape of the 
experimental energy spectrum, being closest to the data in the pion energy range
of 250 - 500 MeV. The models
in general agree with the data within a factor of two.

At the highest detection angles presented here the BUU agrees with the data quite well.
The INCL++ underestimates the data by about a factor two or more 
in the low and high energy ranges.

\begin{figure}[!h]
\includegraphics[width=0.5\textwidth] {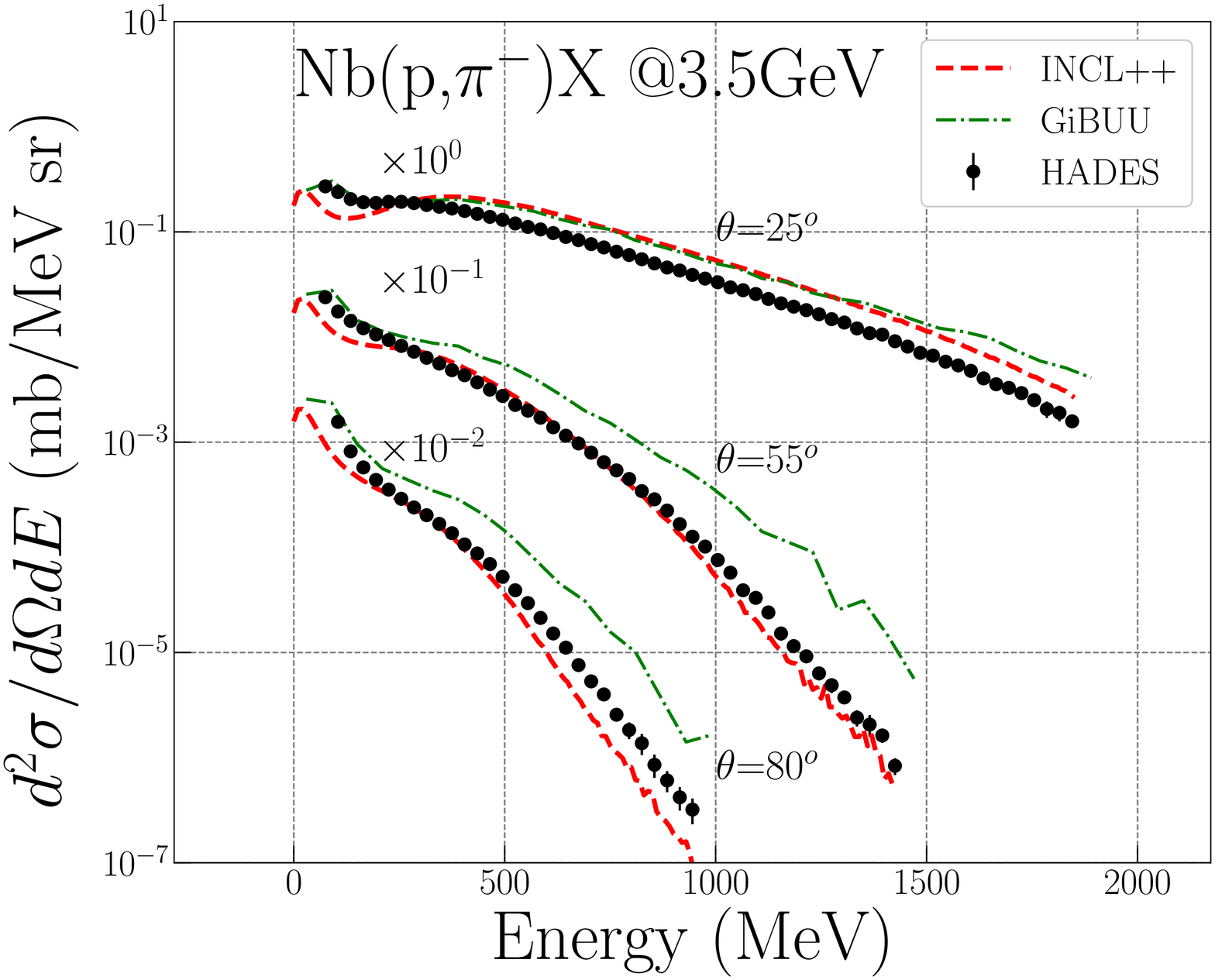}%
\caption{\label{pim_m_3a} The same as in fig. \ref{p_m_3a} but for
$\pi^{-}$.
}
\end{figure}

From the comparison shown in fig. \ref{pim_m_3a} it can be concluded that 
GiBUU overestimates the data for all emission angles and energies of
$\pi^{-}$. The discrepancies, except at the smallest energies, 
are at least a factor two. For larger emission angles 
INCL++ describes the data quite successfully,
however, it fails for the
lowest and highest energies of the detected pions. 

\subsection{\label{composite_part} Composite nuclear particles}

Suitable mechanisms for the formation of composite nuclear
products within an intranuclear cascade are not known. There are various
hypotheses, which use more or less theoretically justified
assumptions (see e.g. \cite{LET02A,HODGSON20031,Iwamoto_2008,Wei_2014,PhysRevC.91.011602,Typel_PhysRevC_81_015803}). 

For low and medium energy pA reactions the most popular hypotheses
are based on coalescence as the 
origin of composite particles during the pre-thermalization phase, 
at least for H and He isotopes. 
Unfortunately, among the tested models 
only INCL++ contains the mechanism of the so-called surface
coalescence, which permits a dynamical construction of composite
particles of masses A $\le$ 8. They can be emitted according to
the conditions defined by the values of the binding energies and the
height of the Coulomb barrier. 

In kinetic transport models the creation of composite particles 
is modeled as well by coalescence in the final state.  
This is done by applying conditions 
on the mutual distances of nucleons in phase space  
after they are emitted from the target nucleus 
or after the freeze-out in heavy-ion collisions 
\cite{Kapusta_PhysRevC.21.1301,Monreal_PhysRevC.60.031901,CHEN2003809,Sharma_N_PhysRevC.98.014914}. 
These methods do not contribute to the 
dynamics of intranuclear cascade and are therefore not considered in this work. 
Promising approaches are under development 
and are partially included e.g., in the newest versions 
of the PHQMD \cite{PHQMD_PhysRevC.101.044905} and SMASH \cite{SMASH_Mohs} 
models but not in the used GiBUU version. 
We hope that the presented data help to further scrutinize those models. 

\begin{figure}[!h]
\includegraphics[width=0.5\textwidth] {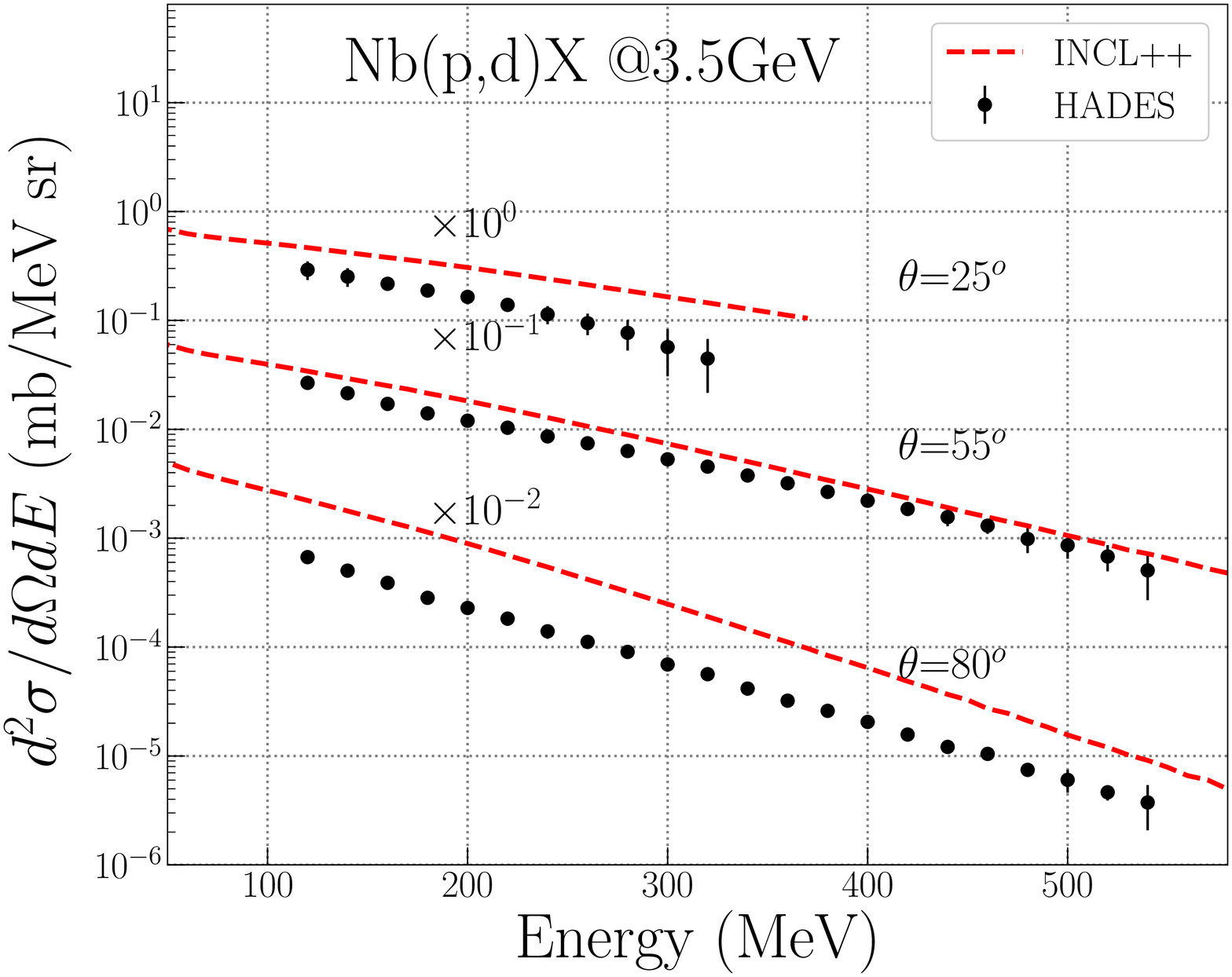}%
\caption{\label{d_m_3a}
%The same in fig. \ref{p_m_3a} but for deuterons.
Double differential cross sections of deuterons measured by HADES in
p (3.5 GeV) + $^{93}$Nb reactions (full circles). 
Cross sections are shown for three polar laboratory emission
angles of $\theta$ = 25$^{\circ}$, $\theta$ = 55$^{\circ}$ (multiplied
by factor 10$^{-1}$) and $\theta$ = 80$^{\circ}$ (multiplied by factor
10$^{-2}$). The experimental distributions are compared with the
results of INCL++ (dashed lines). The constant
normalization uncertainty of 15\% of the experimental data is not shown.}
\end{figure}

\begin{figure}[!h]
\includegraphics[width=0.5\textwidth] {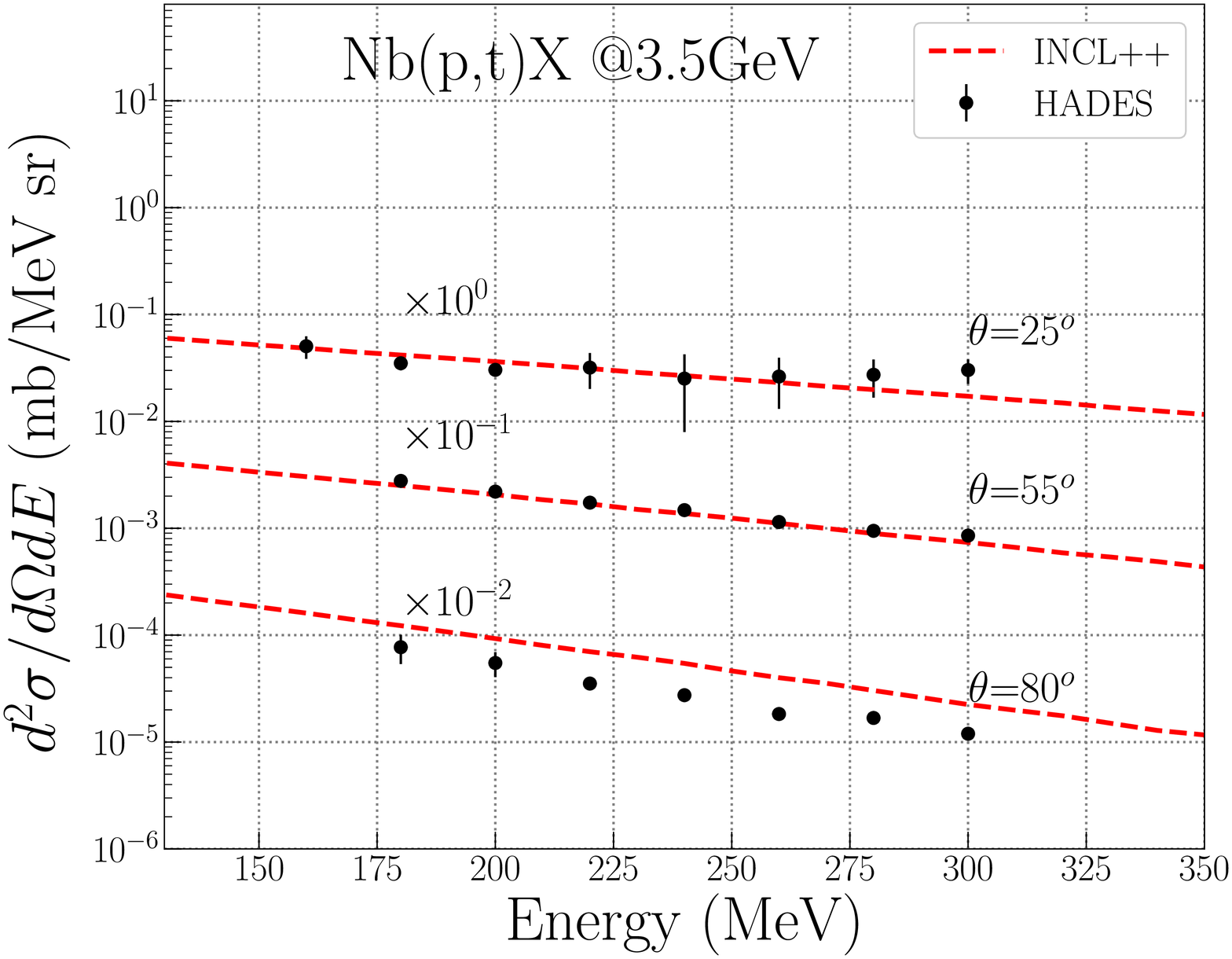}%
\caption{\label{t_m_3a} The same as in fig. \ref{d_m_3a} but for
tritons.
}
\end{figure}

In figs. \ref{d_m_3a} and \ref{t_m_3a}, the HADES results 
for deuterons and tritons, respectively,
are confronted with the predictions by the INCL++  model.
The surface coalescence model implemented in INCL++ generally 
overestimates the production of deuterons. 
For the forward emission angles of deuterons the slope
of the theoretical curve is less steep than for the experimental
distribution. With increasing energy of the emitted deuterons the
discrepancy increases, reaching factors of $\sim$2.5 at energies
$>$ 300 MeV.
For higher emission angles the slopes of the theoretical
distributions are the same as for the experimental ones.
For $\theta$ = 55$^{\circ}$ the magnitude of cross sections 
differs by less than a factor two, whereas for $\theta$ = 80$^{\circ}$ 
it reaches already a factor of about four.

The limitations of the particle identification method based 
on the measurement of the $dE/dx$ vs. $momentum$
dependences have a strong effect on the accessible 
energy range for the triton cross sections. 
Nevertheless, as for the other hydrogen isotopes it was
possible to obtain distributions which extend in energy
beyond existing earlier experimental data.
The HADES results for triton are shown in fig. \ref{t_m_3a}.
Astonishingly, the INCL++ model works better for the triton
differential production cross sections than for deuterons.
For tritons emitted at $\theta$ = 55 $^{\circ}$ the model curve
agrees well with the experimental one over the whole 
measured energy range. Also for lower emission angles, the
agreement of model and experiment is quite good with a slight
underestimate of the experimental cross section at higher energies.
Only at the highest angles measured by HADES, the model
overestimates the experiment by a factor 2, whereas the slopes of
experimental and model distributions are in good agreement.

\section{\label{discussion} Discussion of results}

In the previous section the comparison of the experimental spectra
measured at three different angles with predictions of two models 
are discussed.  It was found that the main
properties of the data, such as a smooth decrease of the cross sections
with increasing scattering angle and energy of the emitted
particles, are reproduced by both models, 
while in detail there are discrepancies. 
It is evident that for
a closer investigation of the quality of the data reproduction by the
models, additional measures useful for comparing data and models have to be involved. 
In order to perform a quantitative assessment 
of the examined models the
method developed in \cite{SHA17A,SIN18A} is applied.
The application of so-called deviation factors is commonly used for 
the quantization of the validity of theoretical models, where 
conventional tools like the $\chi^2$-square test are not 
adequate for the studied problem. Various approaches 
were proposed for this purpose (see e.g. \cite{MICHEL1997153,KURENKOV1999541,Konobeyev}).
A critical analysis of the applicability of various deviation factors 
to the cross section distributions typical for pre-equilibrium component 
of spallation reactions has been performed in \cite{S_Sharma_thesis}.
For example, the popular $H$-deviation factor 
which, under some conditions, is equivalent to the $\chi^2$-square 
test is not applicable for the cases where both, the variation 
of the distributions and their uncertainties, are large.
It was shown in \cite{S_Sharma_thesis,SHA16A,SHA17A} that for a validation 
of cross section distributions in spallation physics  
the $A$-deviation factor is optimal. Hence, it is utilized here.  

The deviation between two discrete distributions 
of cross sections can be quantified by a number between 0 and 1 
defined by 
\begin{equation}
A \equiv \frac{1}{N}   \sum_{i=1}^{N}  \frac{\left |\sigma^{exp}_{i}
- \sigma^{th}_{i}\right |}{\sigma^{exp}_{i} + \sigma^{th}_{i}},
\label{A_1}
\end{equation}
where $\sigma^{exp}_{i}$ and $\sigma^{th}_{i}$ are the values of
the experimental and theoretical cross sections in the $i$-th histogram bin, 
respectively, and $N$ is the total number of histogram bins.

In order to give a consistent comparison of the models and the present
data as a function of kinetic energy $E$ of the reaction product,
as well as its laboratory emission angle $\theta$, the
$A$-quantity was calculated for each bin of the two dimensional histograms 
$\theta$ vs. $E$, i.e. without averaging over several bins 
(as done in eq. (\ref{A_1})):
\begin{equation}
A_{i} \equiv  \frac{\left | \left(d^{2}\sigma/d\Omega dE \right )^{exp}_{i} 
- (d^{2}\sigma/d\Omega dE)^{th}_{i}\right|}
{(d^{2}\sigma/d\Omega dE)^{exp}_{i} + (d^{2}\sigma/d\Omega dE)^{th}_{i}}
\label{A_2}
\end{equation}
(here ($d^{2}\sigma/d\Omega dE)^{exp}_{i}$ and ($d^{2}\sigma/d\Omega dE)^{th}_{i}$ 
are the values of experimental and theoretical differential cross sections 
for a given bin $i$ in the 2D histogram of emission angle vs. energy).

The quantity $A$ vanishes ($A=0$) for an ideal agreement, 
whereas its value increases with the deviation between
the experimental and theoretical  cross sections.  Its highest
possible value is equal to unity ($A=1$), if one of the compared
cross sections vanishes or if its value becomes infinite. In spite
of such an asymmetrical behavior of $A$ for large deviations
between the compared cross sections, it has a very appealing property
for small differences.  Namely, that the $A$ 
values may be interpreted as the proxy of the half of the 
relative distance between the data and theoretical cross sections.
For example, when $A$ = 0.1 the average relative distance between
experimental and theoretical cross sections is close to 20$\%$, and
for $A$ = 0.2 the average deviation of the cross sections is close to
40$\%$.  

The total statistical and systematic uncertainties of the 
data presented in this work
is usually below 20\%. Thus, the resulting uncertainty of $A$ 
remains below the value of 0.1.

\begin{figure*}[!htb]
\includegraphics[width=0.85\textwidth] {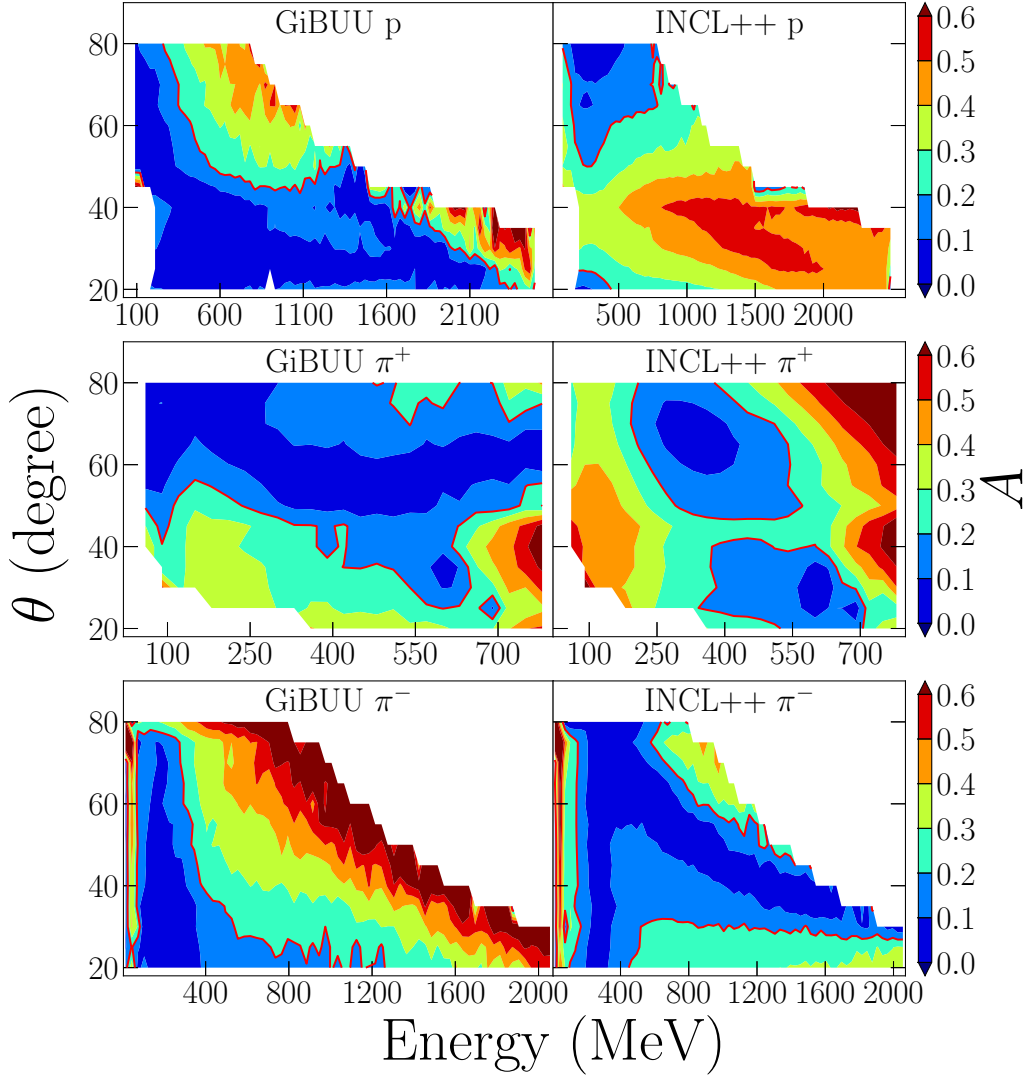}%
\caption{\label{Afp_pip_pim_2D} Polar laboratory emission angle $\theta$ 
and kinetic energy $E$ dependent distributions of the quantity $A$ 
for p (upper row),
$\pi^{+}$ (middle row) and $\pi^{-}$ (lower row). It is calculated
according to eq. (\ref{A_2}) for the GiBUU (left
column) and INCL++ (right column) models in comparison with 
the experimental values of production cross sections 
measured with HADES.
The kinematic regions, where data and model differ by less than 40\% 
(i.e. $A$ $\le$ 0.2), are plotted with blue and dark-blue colors 
and are surrounded by the red contour line.
The uncertainty of $A$ presented in this figure is below a value of 0.1.
Note the different energy regions for the figures in different rows.}
\end{figure*}

Figure \ref{Afp_pip_pim_2D} presents the contour plots for $A$ 
as a function of the 
particle's kinetic energy $E$ and polar emission angle
$\theta$. 
It confirms to a large extent
the conclusions derived from the qualitative analysis given in the previous
section. In the upper set of the two panels of fig.~\ref{Afp_pip_pim_2D}, 
representing proton data, the dark-blue and blue areas, 
where the discrepancy between data and model is smaller than 40\%, 
are clearly larger for GiBUU than 
for INCL++. This proves that GiBUU results in a satisfactory
agreement for a significantly larger number of angles and energies
than INCL++.  Moreover, the region of small values of 
$A$ of INCL++ corresponds only to large angles
(larger than 50$^{\circ}$) and relatively small energies (smaller
than 1000 MeV), whereas GiBUU reproduces well the data at small
angles (smaller than 50$^{\circ}$) and at much broader range of
energies for these angles, i.e. up to $E$ = 1500 - 2000 MeV (depending
on the scattering angle) but also at angles larger than 50$^{\circ}$, 
albeit at the cost of smaller energy range (up to 400 - 500
MeV).

The middle two panels of fig. \ref{Afp_pip_pim_2D}, 
correspond to positively charged pions. It is evident 
that the regions of angles and energies well described by GiBUU 
cover angles larger than 50$^{\circ}$, but the full range of pion
energies (up to 800 MeV). The situation is different for the INCL++
model which describes satisfactorily well the data for almost all
angles, however, in a limited range of energies from about 200 MeV to
500 MeV (at angles larger than 50$^{\circ}$) and from about 300 MeV to
700 MeV (at angles smaller than 40$^{\circ}$).

The lowest two panels of fig. \ref{Afp_pip_pim_2D} represent
the analysis of negatively charged pions $\pi^{-}$ with the GiBUU 
and the INCL++ models (from the left to the right panel of the figure). 
It is clear that each of the models describes well different parts 
of the energy-angle plane: 
the GiBUU model reproduces the smallest fraction of the data.  
These data were reproduced for all angles, however, 
only for a small energy range, which decreases further 
when the emission angle increases. 
At the smallest angles (20$^{\circ}$ - 25$^{\circ}$), 
these energies belong to the interval 
from 100 MeV to 600 - 700 MeV, whereas at largest angles 
lie in  the interval from about 100 MeV to 250 MeV. 
INCL++ describes well most of the $\pi^{-}$ data. 
These data are mainly located at angles larger than 30$^{\circ}$, and 
cover a broad range of energies from 200 MeV 
to 2000 MeV at 30$^{\circ}$. This energy range 
decreases with increasing angle to the range 200 MeV to 650 MeV 
at the largest angle of 80$^{\circ}$.

The data for the emission of deuterons and tritons 
can be compared only with 
predictions of the INCL++ model because GiBUU does not
include a mechanism for cluster formation.

\begin{figure*}[!htb]
\includegraphics[width=0.95\textwidth] {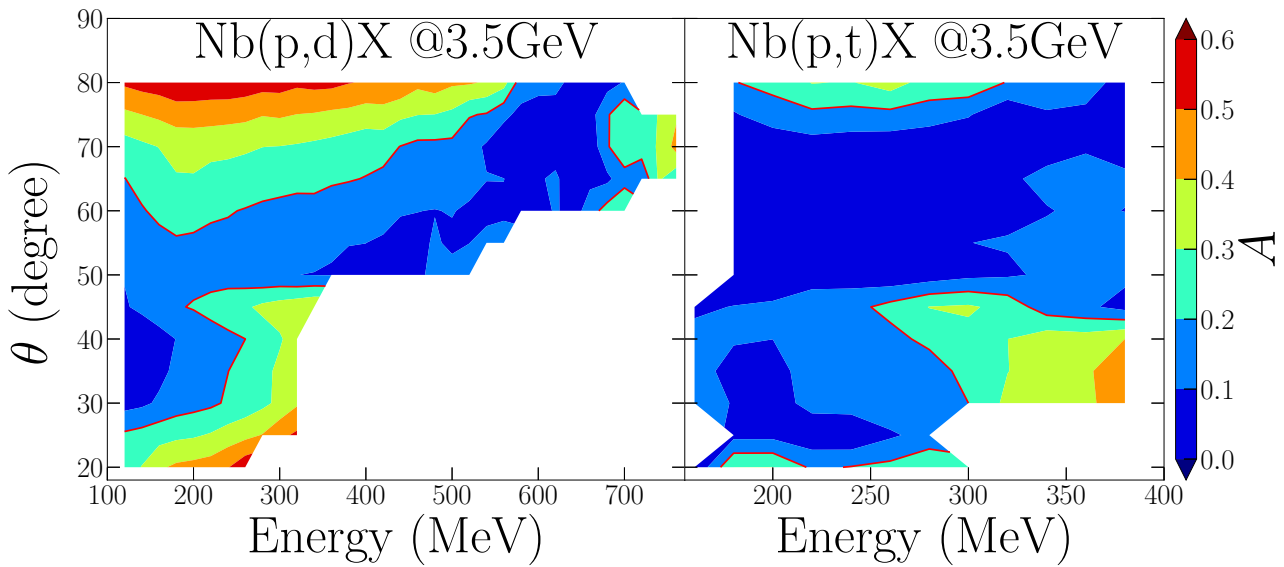}%
\caption{\label{Afd_t_theta_E} Dependence of the $A$ quantity 
on the deuteron (left panel) and triton (right panel) polar emission angle
$\theta$ and kinetic energy $E$. The quantity $A$ is calculated according
to eq. (\ref{A_2}) to compare experimental double
differential cross sections measured by HADES with the results of the 
INCL++ model. 
The kinematic regions where data and model differ by less than 40\% 
(i.e. $A$ $\le$ 0.2) are plotted with blue and dark-blue colors,  
and are surrounded by the red contour line. 
The uncertainty of $A$ presented in this figure is below 0.1.
}
\end{figure*}

The $A$ values for deuteron and triton production are presented
in fig. \ref{Afd_t_theta_E} in the form of two-dimensional maps showing
the dependence on the isotope energy and polar emission angle. 
The behavior of $A$ for deuterons indicates that 
discrepancies smaller than 40\% 
may be expected for the full energy range only for
angles about 50$^{\circ}$ - 60$^{\circ}$, while for tritons such an agreement is
obtained for all angles larger than about 40$^{\circ}$ and additionally
for energies smaller than approximately 270 MeV at angles smaller than
40$^{\circ}$.

The quantity $A$ allows 
to characterize 
the predictive power of the used models. 
This is achieved by determining the fraction of studied
two-dimensional space of energy-emission angle for which $A$ 
values are smaller or equal to 0.2. The ratio of the number 
of these bins to the number of all observed bins can be used as a numerical
measure of the predictive power (PP) of a given model for selected set
of observed particles. 
The corresponding numbers for the INCL++ and GiBUU models 
are given in table \ref{Table1}. 

\begin{table*}%[H] add [H] placement to break table across pages
%\begin{center}
\caption{\label{Table1} Measure of the predictive power (PP) of the GiBUU  
and the INCL++ models for double differential cross sections of
p, $\pi^{+}$, $\pi^{-}$, d and t as measured by HADES. PP is
equal to the fraction of the area (in [\%]) of the $\theta$ vs. $E$
distributions presented in figs. \ref{Afp_pip_pim_2D} and
\ref{Afd_t_theta_E}, where the agreement of the models and
the experimental spectra of HADES is better than 20\% ($A$ $<$ 0.1) 
or 
better than 40\% ($A$ $<$ 0.2). PP for the simulation 
of the intranucelar cascade 
is given for the sum of p, $\pi^{+}$ and
$\pi^{-}$ ejectiles. The correctness of the reproduction of composite
particle production is calculated for the sum of d and t (only
for INCL++). The overall agreement of the INCL++ model with the data
for all particles detected in HADES is given for the sum of them.
The numbers corresponding to several emitted types of particles were calculated 
as the percentage of "good" bins for given set of particles  among all bins 
corresponding to this set of particles.
}
\begin{ruledtabular}
\begin{center}
\begin{tabular}{| p{3.0cm}|| p{1.5cm} | p{1.5cm} || p{1.5cm} | p{1.5cm} |}
\bf{Ejectile} & \multicolumn{2}{c||}{\bf{GiBUU}} & \multicolumn{2}{c|}{\bf{INCL++}}
\\
\hline & $A$ $<$ 0.1 & $A$ $<$ 0.2 & $A$ $<$ 0.1 & $A$ $<$ 0.2
\\
\hline \hline $p$ & 39\% & 65\% & 5\% & 19\%
\\
\hline $\pi^{+}$ & 28\% & 58\% & 27\% & 35\%
\\
\hline $\pi^{-}$ & 12\% & 28\% & 33\% & 63\%
\\
\hline d & & & 18\% & 52\%
\\
\hline t & & & 43\% & 76\%
\\
\hline \hline d+t & & & 27\% & 60\% 
\\
\hline \hline p+$\pi^{+}+\pi^{-}$ & 26\% & 49\% & 16\% & 39\% 
\\
\hline \hline p+$\pi^{+}+\pi^{-}$+d+t & & & 19\% & 43\% 
\end{tabular}
\end{center}
\end{ruledtabular}
\end{table*}

\section{\label{summary} Summary}

Experimental distributions of double differential cross
sections $d^2\sigma/d\Omega dE$ for p,  d,  t, $\pi^{+}$ and
$\pi^{-}$ production in p (3.5 GeV) + Nb reactions have been
extracted from data collected by a HADES experiment. 
The quality of the obtained 
cross sections has been verified by comparisons with other
available experimental results published in the literature. 

The data provided in this paper are measured in the angular range 
from 20$^{\circ}$ to 80$^{\circ}$ of the laboratory polar emission angle
$\theta$. Due to the high acceptance and the magnetic field
of HADES the measured cross sections for almost all
detected species and detection angles exceed the energy ranges of 
data available up to now in the literature. The most significant
extension of the measured kinematical region was obtained for proton
data.

The obtained cross sections have been compared to the
results of two reaction models, namely INCL++ and GiBUU.
The comparison of the shapes and magnitudes of the experimental
and theoretical cross sections 
show that, in general, these models are able to
reproduce the data within a factor of 2. 
Unfortunately, the correctness of the data
description varies 
in a non-trivial way 
for each model with the type of
produced particle, its energy and emission angle.

A quantitative comparison between data and model predictions is also done 
using the $A$-quantity, commonly employed 
for a comparison of spallative emission of slow particles with statistical models
\cite{SHA17A,SIN18A}. 
This comparison shows that a better description 
of the angular and
energy dependent distributions of protons and charged pions is provided by 
the GiBUU model, whereas the accuracy of the description 
by INCL++ is by a factor of $\sim$1.5 worse. 

The origin of energetic nuclear clusters (d and t) can be inferred from a 
comparison of the double differential cross sections for the production
of deuteron and triton registered in HADES with the results of the INCL++
model. The surface coalescence mechanism implemented in the INCL++ model as
the source of light nuclei reproduces quite well the
distributions of tritons. The modelled deuteron spectra
overestimate however the data in the whole energy and angular range.

At the current stage of the theoretical examination of the dynamics
of intranuclear cascades and the phenomena responsible for 
the clustering of nuclear matter in thermal pre\-equilibrium, 
a clear conclusion about the validity of proposed scenarios 
cannot be drawn.  
The precision of the models is still 
not sufficient in order to perform a detailed verification of their
features by a comparison to the experimental data. 
For example, the GiBUU model describes the p and $\pi^{+}$ spectra better
than INCL++, but for $\pi^{-}$ its results are worse than 
for the INCL++ model. 
GiBUU suffers as well from the lack of composite
particle production. Introducing such a mechanism to this model would
certainly modify also the theoretical p distributions. 
Thus, any judgments in favor of one of the two 
models tested here would be premature.

\begin{acknowledgments}
The collaboration gratefully acknowledges the support by 
SIP JUC Cracow, Cracow (Poland), 2017/26/M/ST2/00600; TU Darmstadt, Darmstadt (Germany), VH-NG-823, DFG GRK 2128, DFG CRC-TR 211, BMBF:05P18RDFC1, HFHF, ELEMENTS 500/10.006, EMMI at GSI Darmstad; Goethe-University, Frankfurt (Germany), BMBF:05P12RFGHJ, GSI F\&E, HFHF, ELEMENTS 500/10.006, HIC for FAIR (LOEWE), EMMI at GSI Darmstadt; JLU Giessen, Giessen (Germany), BMBF:05P12RGGHM; IJCLab Orsay, Orsay (France), 
CNRS/IN2P3; NPI CAS, Rez, Rez (Czech Republic), MSMT LTT17003, MSMT LM2018112, MSMT OP VVV CZ.02.1.01/0.0/0.0/18\_046/0016066; WUT Warszawa (Poland) No: 2020/38/E/ST2/00019 (NCN), IDUB-POB-FWEiTE-3; \\
The following colleagues from Russian institutes did contribute to the results presented in this publication but are not listed as authors following the decision of the HADES Collaboration Board on March 23, 2022: G. Agakishiev, A. Belyaev, O. Fateev, A. Ierusalimov, V. Ladygin, T. Vasiliev, M. Golubeva, F. Guber, A. Ivashkin, T. Karavicheva, A. Kurepin, A. Reshetin, A. Sadovsky.
\end{acknowledgments}

\pagebreak


\bibliography{hades_spallation_v6_3.bib}

\end{document}